\documentclass[12pt, a4paper]{article}
\usepackage[utf8]{inputenc}
\usepackage{afterpage}
\usepackage{algorithm}
\usepackage{algorithmicx}
\usepackage{algpseudocode}
\usepackage{authblk}
\usepackage{amsmath}
\usepackage[title]{appendix}
\usepackage{amsfonts}
\usepackage{booktabs} 
\usepackage[
    style=numeric-comp,
    sorting=none,
    maxbibnames=20
]{biblatex}
\renewcommand*{\cite}{\supercite}
\usepackage[font=small,labelfont=bf, hypcap=false]{caption} 
\usepackage{chngcntr}
\usepackage{comment} 
\usepackage{csquotes} 
\usepackage{csvsimple} 
\usepackage{diagbox}
\usepackage{enumitem}
\usepackage{eqnarray}
\usepackage{float} 
\usepackage[T1]{fontenc} 
\usepackage{geometry}
\usepackage{graphicx}
\usepackage[colorlinks=true, allcolors=blue]{hyperref}
\usepackage[none]{hyphenat} 
\usepackage[right]{lineno}  
\modulolinenumbers[5]
\usepackage{listings}
\usepackage{lipsum}
\usepackage{longtable}
\usepackage{lscape}
\usepackage{mathrsfs}
\usepackage{multirow}
\usepackage{orcidlink}
\usepackage{pdflscape}
\usepackage[]{ragged2e}
\usepackage{subcaption}
\usepackage{setspace} 
\usepackage{threeparttable} 
\usepackage{tabularray}
\usepackage{titlesec} 
\usepackage[english]{babel}
\usepackage{hyphenat}
\usepackage{microtype}
\usepackage{graphicx} 

\usepackage{color}
\newcommand{\revtwo}[1]{\textcolor{black}{#1}}

\titleformat{\section} 
{\normalfont\Large\bfseries}{\thesection.}{1em}{}

\linenumbers 
\rightlinenumbers 



\title{\textbf{Habit learning is associated with efficiently controlled network dynamics in naive macaque monkeys}}
\author[1,\textsuperscript{\textdagger}]{\small Julia K. Brynildsen}
\author[1,2,\textsuperscript{\textdagger}]{\small Panagiotis Fotiadis}
\author[1,3,4,\textsuperscript{\textdagger}]{\small Karol P. Szymula} 
\author[1,5]{\small Jason Z. Kim}
\author[6]{\small Fabio Pasqualetti}
\author[7,8]{\small Ann M. Graybiel}
\author[9,$\ddag$,*]{\small Theresa M. Desrochers}
\author[1,10,11,12,13,14,$\ddag$,*]{\small Dani S. Bassett}
\affil[1]{\small{Department of Bioengineering, School of Engineering \& Applied Science, University of Pennsylvania, Philadelphia, PA 19104 USA}}
\affil[2]{\small{Department of Neuroscience, Perelman School of Medicine, University of Pennsylvania, Philadelphia, PA 19104, USA}}
\affil[3]{\small{Department of Biomedical Engineering, University of Rochester, Rochester, NY 14642 USA}}
\affil[4]{\small{Medical Scientist Training Program, University of Rochester School of Medicine and Dentistry, Rochester, New York, USA}}
\affil[5]{\small{Department of Physics, Cornell University, Ithaca, NY, 14853, USA}}
\affil[6]{\small{Department of Mechanical Engineering, University of California, Riverside, CA 92521 USA}}
\affil[7]{\small{McGovern Institute for Brain Research, Massachusetts Institute of Technology, Cambridge, MA 02139 USA}}
\affil[8]{\small{Department of Brain and Cognitive Sciences, Massachusetts Institute of Technology, Cambridge, MA 02139 USA}}
\affil[9]{\small{Department of Neuroscience, Department of Psychiatry and Human Behavior, Robert J. and Nancy D. Carney Institute for Brain Science, Brown University, Providence RI 02912 USA}}
\affil[10]{\small{Department of Electrical \& Systems Engineering, School of Engineering \& Applied Science, University of Pennsylvania, Philadelphia, PA 19104 USA}}
\affil[11]{\small{Department of Physics \& Astronomy, College of Arts \& Sciences, University of Pennsylvania, Philadelphia, PA 19104 USA}}
\affil[12]{\small{Department of Neurology, Perelman School of Medicine, University of Pennsylvania, Philadelphia, PA 19104 USA}}
\affil[13]{\small{Department of Psychiatry, Perelman School of Medicine, University of Pennsylvania, Philadelphia, PA 19104 USA}}
\affil[14]{\small{Santa Fe Institute, Santa Fe, NM 87501 USA}}
\affil[ \textsuperscript{\textdagger}]{These three authors contributed equally.}
\affil[$\ddag$]{\small{These two authors contributed equally.}}
\affil[*]{Corresponding authors: 
  \href{mailto:theresa_desrochers@brown.edu}{theresa\_desrochers@brown.edu}; 
  \href{mailto:dsb@seas.upenn.edu}{dsb@seas.upenn.edu}} 
\date{}

\begin{document}
\maketitle
\clearpage  
\begin{abstract}
\noindent{Primates utilize distributed neural circuits to learn habits in uncertain environments, but the underlying mechanisms remain poorly understood. We propose a formal theory of network energetics explaining how brain states influence sequential behavior. We test our theory on multi-unit recordings from the caudate nucleus and cortical regions of macaques performing a motor habit task. The theory predicts the energy required to transition between brain states represented by trial-specific firing rates across channels, assuming activity spreads through effective connections. We hypothesized that habit formation would correlate with lower control energy. Consistent with this, we observed smaller energy requirements for transitions between similar saccade patterns and those of intermediate complexity, and sessions exploiting fewer patterns. Simulations ruled out confounds from neurons’ directional tuning. Finally, virtual lesioning demonstrated robustness of observed relationships between control energy and behavior. This work paves the way for examining how behavior arises from changing activity in distributed circuitry.}
\end{abstract}

\newpage
\newpage
\section*{Introduction}

In a complex ever-changing environment, both humans and non-human primates survive by learning to balance the need to gather new knowledge with the utilization of existing knowledge\cite{costa2019subcortical,ebitz2018exploration}. The formation of habits can be viewed as a natural consequence of locally optimal trade-offs between exploration and exploitation\cite{desrochers2010optimal}. The underlying cognitive processes may follow reinforcement learning algorithms\cite{kaelbling1996reinforcement}, in which the sampling of actions and the uncertainty of their outcomes inform decisions regarding exploration of new actions or exploitation of old ones\cite{gershman2018deconstrucing}. The brain mechanisms supporting such processes engage a distributed set of regions spanning the caudate nucleus associated with repetitive and stereotyped actions\cite{desrochers2015habit}, the ventral striatum and amygdala associated with reward and motivation\cite{costa2019subcortical}, and the prefrontal cortex associated with cognitive control\cite{ebitz2018exploration}.

Yet, precisely how this constellation of brain regions supports the computations necessary for habits to emerge remains far from understood. Recent efforts suggest that network approaches\cite{mitchell2011complexity} provide useful tools for understanding how cognitive processes arise from interacting brain regions\cite{rosenberg2020functional}. From intelligence to cognitive control, and from motivation to learning, disparate circuits are engaged that allow coordinated information processing and transmission\cite{barbey2018network,girn2019linking,bassett2017network}. The study of circuit engagement and function can be formalized in the language of network science\cite{bassett2018on}, and initial evidence suggests that individual differences in the pattern of inter-regional interactions track individual differences in exploratory behaviors and decision-making\cite{kao2019functional}, plasticity\cite{gallen2019brain}, reinforcement learning\cite{gerraty2018dynamic}, and skill learning\cite{bassett2015learning}. Although network approaches manifest striking face validity, the level of explanation has thus far largely neglected the role of neural connectivity in shaping and constraining patterns of neural activity associated with cognitive and behavioral processes\cite{bertolero2019on}.

Here we address this challenge by building upon and extending emerging work in the field of network control theory\cite{motter2015network,pasquletti2014controllability,liu2011controllability}. In the context of neural systems, the approach defines a state of the network to be the vector of regional (or cellular) activation. The theory then posits that the sequence of states is constrained by the energy required to transmute one state into another allowing activity to spread solely through known inter-regional links\cite{kim2019linear}. In addition to predicting the effects of exogenous control signals such as electrical stimulation\cite{stiso2019white}, network control theory has also proven useful in accounting for the intrinsic capacity for cognitive control\cite{cornblath2019sex,zhou2023} and the contribution of single neurons to large-scale behaviors\cite{yan2017network}. We extend the approach in two ways. First, prior studies stipulated that activity could only flow along known structural links between regions; here, we instead allow inter-regional links to reflect effective connections\cite{friston2011functional}, which represent information flow in the brain during learning\cite{dima2015neuroticism,buchel1999predictive}. We assume that effective connectivity remains constant over the course of the study and attempt to capture learning-related changes through changes in endogenous inputs. Second, prior studies estimated the energy of brain state transitions independently from behavior; here, we instead explicitly posit (and validate) the notion that low energy state transitions characterize processes that are less cognitively demanding, as do their associated behaviors\cite{braun2021brain}. 

We evaluate the theory in the context of multi-unit recordings spanning the caudate nucleus and prefrontal cortex of two macaque monkeys as they engage in 60-180 sessions of task performance inducing motor habits in the form of saccadic patterns\cite{desrochers2010optimal}. In each trial, the monkey was presented with a 3$\times$3 grid of green dots and was expected to freely traverse its gaze across the 9 dots (\hyperref[fig1]{Figure 1}). At some variable point during the trial, one of these 9 dots was randomly “baited” unbeknownst to the monkey (the baited target was different across each trial and indistinguishable from the other dots). Once the monkey’s gaze entered the baited dot area, the trial ended and a reward was given. Due to the nature of the task, the optimal behavior would thus be for the monkey to perform a sequence of saccadic movements that spanned all dots in a minimal amount of time. We define a brain state to be a vector of firing rates across recording channels within a trial. Further, we construct a neural network whose nodes are channels and whose edges reflect the strength of effective connectivity derived from neural activity time-series; we represent the network as a weighted directed adjacency matrix. 

Habits are characterized by repetitive, automatized patterns of behavior. Our primary hypothesis is that pairwise differences in sequential behaviors, as observed across individual trials during habit formation, will be associated with changes in energy requirements of the accompanying neural state transitions. To interpret behavior, we represent saccade patterns as graphs that can be decomposed into 1-D time-series (\hyperref[fig1]{Figure 2A}), whose shape can be studied and whose complexity can be quantified. We observe that preferred saccade patterns change as a function of learning, in a manner that is consistent with the formation of habits. Specifically, we observe increases in the degree of similarity between complex saccade patterns and repetition of stereotypical saccade sequences across time. Using network control theory, we compute the minimum control energy required to transition between chronologically ordered trial brain states where a “state” reflects neural activity across a trial, \revtwo{and may thus reflect activity patterns} associated with the behavioral strategy predominantly used within each trial and observe that energy decreases with learning. Finally, we show that the energy of state transitions predicts behavior in three distinct ways: (i) transitioning between more similar saccades patterns requires less energy, (ii) performing saccade patterns of intermediate complexity requires less energy, and (iii) repeatedly exploiting a smaller number of saccade patterns during sessions requires less energy. \revtwo{Together, these findings suggest that lower-energy transitions support behaviors that are less variable and more stereotyped. This pattern is broadly consistent with the principle of maximum entropy, in which energy constraints shape the diversity of accessible behavioral states, as has been observed in other neural and behavioral systems}\cite{savin2017maximum,granot2013stimulus,meshulam2017collective,friston2006free,ortega2013thermodynamics}.

To rule out the influence of inherent directional tuning properties of neurons, we conducted simulations using only the monkeys' saccadic trajectories, excluding recorded firing rates. Unlike the empirical variables, the correlations in the simulated variables were either non-significant or much weaker, indicating that directional tuning did not drive our results. Lastly, we aimed to identify which parts of the brain's connectome contribute most to the relationship between control energy and saccadic performance. We perform virtual lesion analyses, setting connectome edges to zero. Edges whose knockout renders the relationship non-significant are deemed critical. We find that disrupting most edges is required for our observed behavior-energy correlations to become non-significant. Taken together, our work represents a theoretically principled study of habit learning that identifies a relationship between transitions in behavior and the energetics of transitions in neural states. 

\section*{Results}
\begin{figure}[!t]
\centering
\includegraphics[width=\linewidth]{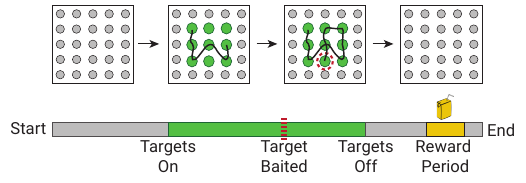}
\caption{\textbf{The Free Scan Task}. A visual depiction of a single trial from the free scan task. The trial begins (\emph{Start}) with a grid of small grey dots all equally sized and spaced. A 3$\times$3 grid of larger green targets replaces the central part of the gray grid (\emph{Targets On}). The monkey is allowed to freely scan the space of green targets. At a variable time unknown to the monkey, a target is baited (\emph{Target Baited}). For visualization purposes in the context of this exposition, the baited target is depicted surrounded by a red dashed circle; note that the outline is not present during the actual task. When the monkey’s gaze enters the baited target, the grid of green targets disappears and is replaced by all gray targets (\emph{Targets Off}). After a brief variable delay, a liquid reward is given to the monkey (\emph{Reward Period}) after which the trial is considered finished (\emph{End}).}
\label{fig1}
\end{figure}

The data analyzed in this work consist of neural recordings and behavioral measurements from two female macaque monkeys, Monkey G (60 recorded sessions; 9,702 task trials) and Monkey Y (180 recorded sessions; 80,664 task trials), while performing a free-view scanning task. All data were previously collected and reported\cite{desrochers2015habit, desrochers2010optimal}. Neural data were recorded from approximately 100 electrodes implanted in the caudate nucleus, Brodmann areas 8 (frontal eye fields), 9 (dorsolateral and medial prefrontal cortex), 13-14 (insula and ventromedial prefrontal cortex), 24-25 (anterior and subgenual cingulate), and 45-46 (\emph{pars triangularis} and middle frontal area). As depicted in \hyperref[fig1]{Figure 1}, \revtwo{at the beginning of each trial} the monkey is shown a 3$\times$3 grid of green targets on a screen and allowed to visually navigate the grid-space freely. At a variable time, one of the targets is baited without the monkey's knowledge. When the monkey's gaze enters the baited target space, the green grid is replaced by smaller gray circles (marking the end of the task trial) and the monkey receives a reward after a short delay \revtwo{On average, each trial lasted $6.2 \pm 1.1$ s (mean $\pm$ SD) for MG and $8.0 \pm 1.3$ s for MY. Within trials, the scan epoch (from scan\_on to scan\_off) lasted $1.9 \pm 1.1$ s for MG and $2.2 \pm 1.2$ s for MY, and the reward delivery epoch (reward\_on to reward\_off) lasted $0.20 \pm 0.002$ s for MG and $0.29 \pm 0.06$ for MY.}

Earlier analyses of this dataset\cite{desrochers2010optimal, desrochers2015habit} indicate that both monkeys showed evidence of learning over repeated trials, with increasingly efficient visits to the targets within each trial and more repetitive saccade patterns (i.e., decreased saccade entropy) across sessions, regardless of the pattern being performed. A reinforcement learning algorithm closely mirrored this behavior. However, while the reward rate—measured as the number of rewards earned per total scan time—remained relatively constant for Monkey G and reached asymptote for the last 18 sessions for Monkey Y, the monkeys' scan patterns continued to evolve well beyond this point\cite{desrochers2010optimal}. These results suggest that scan time alone may not effectively capture habit formation, as changes in scan time did not consistently align with the monkeys’ shifting approach to the task.

\subsection*{Classification of Trial Representative Saccade Patterns}
\begin{figure*}[!t]
\includegraphics[width=\textwidth]{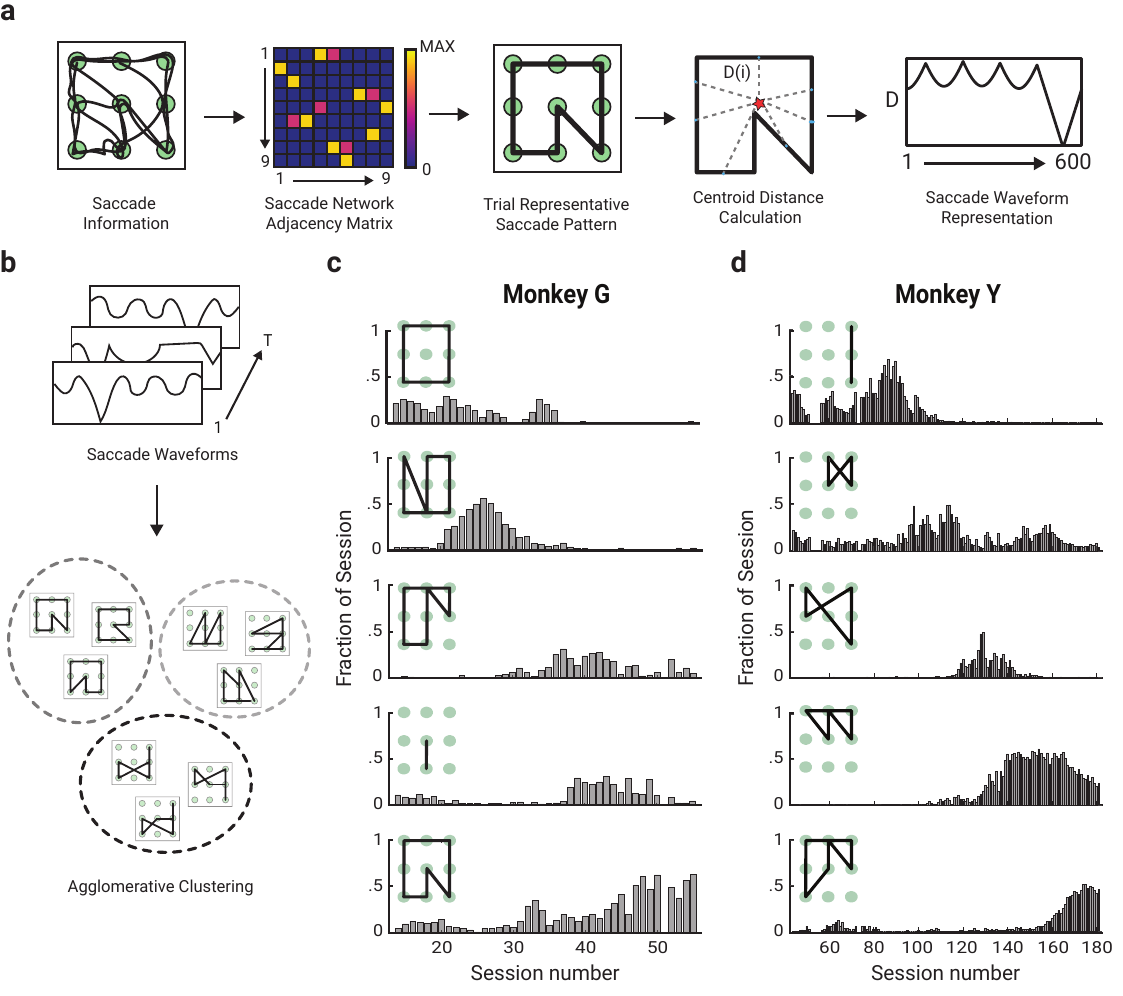}
\caption{\textbf{Classification of Trial Representative Saccade Patterns}. \textbf{(a)} Saccade information in the form of identified saccadic movements during a trial is collectively represented as an adjacency matrix, which in turn encodes a directed and weighted network. A total of nine nodes exist: one for every green target on the task grid. Edge weights are calculated as the number of times that an individual saccade is made from one node to another. The network is converted into a trial representative saccade pattern by identifying the network cycle with the greatest sum of edge weights along its path. Each TRSP is treated as a 2-D polygon in the task grid space consisting of a set of (x,y) points. The saccade waveform is taken to be the vector of Euclidean distances between the polygon centroid and all of its points. A 1-D interpolation is performed to reduce each saccade waveform to 600 values. \textbf{(b)} A dissimilarity matrix is constructed utilizing the saccade waveforms from all observed trials. Each element in the dissimilarity matrix is the Euclidean distance between two saccade waveforms. The dissimilarity matrix is of size $T \times T$ where $T$ is the total number of trials for a given monkey. Agglomerative clustering with a threshold inconsistency coefficient of 0.95 was performed using the dissimilarity matrix to cluster all trial saccade waveforms. For Monkey G, we identified a total of 136 cluster whereas for Monkey Y, we identified a total of 346 clusters. \textbf{(c,d)} The five most prevalent cluster saccade patterns across all sessions are shown for each monkey. The saccade pattern shown is the one which is most similar to all other saccade patterns in the same cluster.}
\label{fig2}
\end{figure*}

Behavioral measurements were analyzed in the form of trial-specific chronological saccade sequences performed by a monkey during the task. We use the phrase \emph{individual saccade} to refer to a rapid eye movement from one target to another on the task grid. The phrase \emph{saccade sequence} then refers to a series of individual saccades that are performed one after another. Direct qualitative or quantitative comparison between the trial-specific saccade sequences is difficult due to variability in trial length and, as a result, the number of saccades per trial. Therefore, we first set out to arrange the list of individual saccades into a format that allows for interpretable comparison between trials. We began by converting each trial's saccade sequence into an adjacency matrix of a directed and weighted graph, $G\left(N,E\right)$, where $N$ is the number of nodes (one for each green grid target) and $E$ is the set of all edges that exist between nodes (\hyperref[fig2]{Figure 2A}). An edge between two nodes exists if a saccade was observed between the two specific grid targets. Furthermore, the weight of each edge is the total number of times the specific saccade is performed during the trial. We refer to this representation of the trial saccades as the \emph{saccade network}. 

Saccade patterns that create \emph{loops} (or sequences that start and end on the same target) \revtwo{in which each target is visited only once before returning to the starting target} are the most effective strategies since they allow for an efficient and organized approach to scanning the 3$\times$3 grid space\cite{desrochers2010optimal}. \revtwo{Indeed, the majority of successful trials for each monkey (75.96\% for MG and 74.64\% for MY) were comprised of looping sequences, underscoring their development as the dominant strategy in successful task performance.} Accordingly, we identified the most observed cyclic saccade pattern in each trial by leveraging the concept of network paths (see \hyperref[subsec:methods:classif]{Methods}). A series of edges that is traversed to move from one node in the network to another is called a path. In a saccade network, a path represents a set of observed saccades performed one after another. If the path's start node is the same as the path's end node then the path is cyclic and the represented saccade sequence is a loop. For each trial, we therefore defined the trial representative saccade pattern (TRSP) to be the cyclic path in the trial saccade network with the greatest sum of edge weights (\hyperref[fig2]{Figure 2A}). Intuitively, the TRSP is the cyclic sequence of saccades that was performed most frequently during a trial; see \textbf{Supplementary Figure 1} for a graphical depiction of TRSP identification.

Methods for computing the similarity between 1-D signals are numerous, easy to implement computationally, and provide simple intuitive understanding. Therefore, to make the comparison between TRSPs as simple as possible we converted each pattern into a 1-D saccade waveform (\hyperref[fig2]{Figure 2A}). This dimensionality reduction step was made possible by representing the identified TRSP as a series of (x, y) points in the space of the task grid and calculating each points’ Euclidean distance from the centroid of all points. We take the similarity between any two trial saccade patterns to be a metric based on the Euclidean distance between the trial saccade waveforms (see \hyperref[subsec:methods:classif]{Methods}). We use this metric to group all the trials into clusters of saccade patterns based on their similarity to each other. Since the exact number of present clusters in the data was not known, the agglomerative clustering algorithm was used to group TRSPs (\hyperref[fig2]{Figure 2B}). This algorithm starts by treating each object (i.e. saccade pattern) as a single cluster and uses an iterative process to merge pairs of objects into clusters until all objects are grouped into one large cluster. The output of the algorithm is a dendrogram (cluster tree) which depicts the order in which objects should be grouped during clustering. \par

In order to capture more natural divisions of our data during clustering, we used the inconsistency coefficient which is a useful metric in agglomerative clustering that compares the height of a link in a cluster tree to heights of all the other links underneath it in the tree. A small coefficient denotes little difference between the objects being grouped together, thereby suggesting that the clustering solution is a good fit to the data. Examining a range of inconsistency coefficient values allowed us to identify an optimal threshold criterion for clustering. Setting a threshold on the inconsistency coefficient during clustering enables the identified groupings to more closely represent the natural divisions found in the data. Using the elbow-method and the average within cluster sum-of-squares from a range of inconsistency coefficient thresholds (0.1-1.5, in intervals of 0.05), we selected an inconsistency coefficient of 0.95 to be the optimal threshold criterion for clustering. As a result, a total of 136 clusters were identified for Monkey G and 346 for Monkey Y; see \textbf{Supplementary Figures 2 and 3} for a full tabulation. In \hyperref[fig2]{Figure 2C}, the five most prominent TRSPs for each monkey as well as their appearance frequency distribution across all sessions are shown. These patterns and their dynamics closely resemble those previously reported\cite{desrochers2010optimal}. Both monkeys demonstrate non-uniform distributions of cluster appearance frequencies across sessions. Each of the main cluster types is acquired, preferentially performed, and dropped at varying time windows throughout the task. \par

\subsection*{Inferring Effective Connectivity}

\begin{figure}[!bp]
\centering
\includegraphics[width=0.6\linewidth]{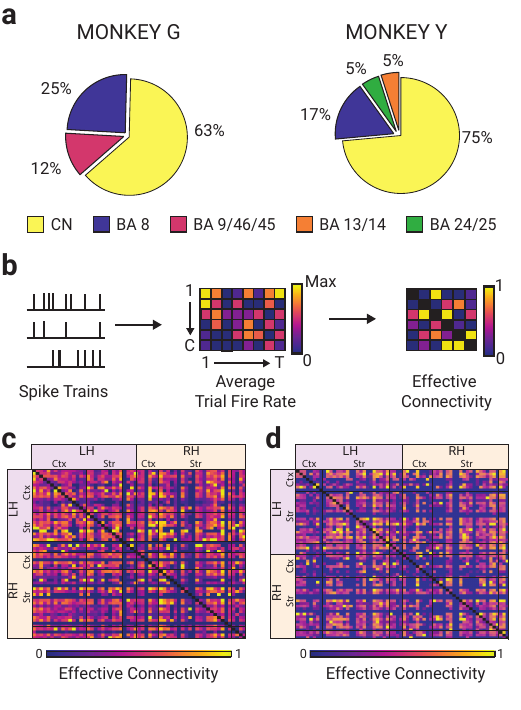}
\caption{\textbf{Inferring Effective Connectivity from Neural Activity}. \textbf{(a)} Summary of channel recording regions for both monkeys. Percentages denote the percent of all available channels which record from a given region across all trials. CN = caudate nucleus; BA 8, 9, 13-14, 24-25, 45-46 = Brodmann areas 8 (frontal eye fields), 9 (dorsolateral and medial prefrontal cortex), 13-14 (insula and ventromedial prefrontal cortex), 24-25 (anterior and subgenual cingulate), and 45-46 (\emph{pars triangularis} and middle frontal area). \textbf{(b)} Spike trains from all channels for a given session were converted into a trial firing rate matrix. The matrix is of size $C\times T$, where $T$ is the number of trials for a session and $C$ is the number of available channels during the session. We used transfer entropy\cite{vicente2011transfer} to estimate the effective connectivity between session channels. \textbf{(c,d)} The overall combined effective connectivity matrices for both monkeys are shown where channels are organized according to their respective hemispheres (LH = Left; RH = Right). Both matrices are individually normalized column-wise by dividing each column by its maximum value; normalization was performed for visualization purposes only. Ctx: Cortex; Str: Striatum.}
\label{fig3}
\end{figure}

After quantitatively characterizing behavior, our next aim was to demonstrate that pairwise differences in sequential saccade patterns during habit formation can be explained by the energy requirements of the accompanying neural state transitions. We approached the problem by using and extending recent advances in network control theory\cite{motter2015network,pasquletti2014controllability,liu2011controllability}. Fundamental to any control energy analysis is knowledge of the network structure and dynamics. Thus, as a first step we extract a network of interactions between the observed regions from available channels (\hyperref[fig3]{Figure 3A}). In both monkeys, more than half of the present channels were associated with the caudate nucleus and recordings from Brodmann area 8 (BA-8) were available in both. Although the anatomical location of each channel was known, no information regarding their anatomical connectivity was available and we therefore turned to alternative inference approaches\cite{friston2011functional}. 

Specifically, we inferred the effective connectivity of the regions using their neural activity (see \hyperref[subsec:methods:ec]{Methods}). For each session, the activity of the available channels was calculated as the firing rate during each individual trial. This set of trial firing rates was used to calculate the transfer entropy\cite{vicente2011transfer} between all pairs of available channels, which provides basic topographical information about the effective connectivity between them (\hyperref[fig3]{Figure 3B}). The effective connectivity matrices for Monkey G and Monkey Y are displayed in (\hyperref[fig3]{Figure 3C, 3D}).  

\subsection*{Assessing the Control Energy Required for Neural State Transitions}

\begin{figure*}[!t]
\includegraphics[width=\textwidth]{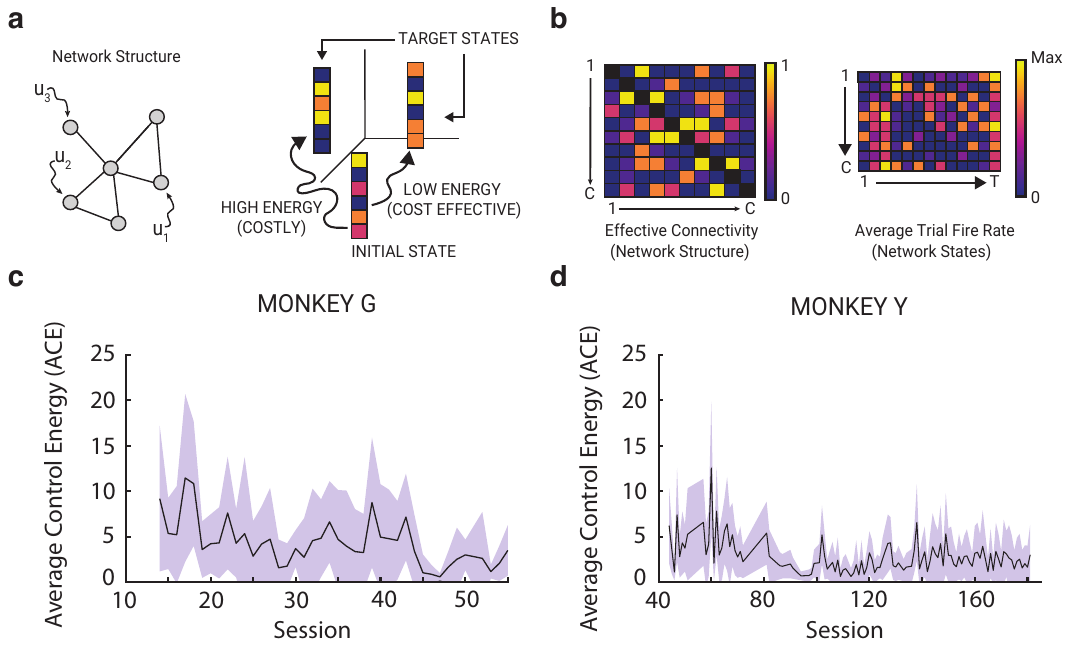}
\caption{\textbf{Estimating the Minimum Control Energy to Transition between Neural States}. \textbf{(a)} Visual depiction of control energy analysis. Given a network, the goal is to identify a set of time-dependent inputs (e.g., $u_1$($t$), $u_2$($t$), and $u_3$($t$)) into network nodes that drives the system from an initial state to a target state in a fixed period of time. A state is a $1 \times N$ vector $x_{t}$ whose elements represent the activity of each of $N$ nodes in the network at some time $t$. The calculation of minimum control energy estimates the energy required to transmute one state into another allowing activity to spread solely through known inter-regional links. The greater the minimum control energy the more costly and hard-to-reach that target state is said to be. \textbf{(b)} For a given session, we model the network of channels as a linear time independent system and compute the minimum control energy required to transition between chronologically ordered trial states. A trial state is taken to be the firing rate of each channel over the course of an entire trial. The topology of the network is defined by a session-based subset of the weighted average matrix, in which nodes vary across sessions depending on channel availability, and effective connections between them are consistent. \textbf{(c-d)} The average minimum control energy (ACE) is calculated across all pairwise trial state transitions of a particular session. ACE dynamics for both monkeys across their respective sessions are shown. Filled boundary areas represent +/- 1 standard deviation.}
\label{fig4}
\end{figure*}

In applying and extending network control theory to understand habit formation, our next step is to use the effective connectivity networks to estimate the energy requirement of neural state transitions. We use the concept of minimum control energy from control theory, which represents the minimum amount of input energy necessary to cause a network to transition from a specific initial state of activity to a specific final state of activity (\hyperref[fig4]{Figure 4A})\cite{stiso2019white}. Intuitively, the more energy a transition requires, the more difficult it is to reach the final state. In the context of neural systems, control energy has previously been found to correlate with cognitive effort\cite{braun2021brain, zhou2023}.

In prior work, minimum control energy has been defined for mechanical and technological systems, or abstract mathematical models. To use the approach here, we must first identify a relevant dynamical model for the considered network process. This model consists of (i) a network state, which we define as the trial firing rates (\hyperref[fig4]{Figure 4B}), (ii) a transition map for the state, which we define as the inferred effective connectivity matrix, and (iii) a set of driver nodes, which include all the nodes in our study (see \hyperref[subsec:methods:controlenergy]{Methods}). Edges within the effective connectivity matrix represent the average of the set of non-zero connections between each pair of channels across all sessions in which both of those channels were available (see \hyperref[subsec:methods:ec]{Methods}). With these variables defined, we then estimate the energy required to transmute one state into another allowing activity to spread solely through effective connections. Specifically, we calculated the average minimum control energy (ACE) theoretically required for the observed trial-to-trial state transitions within each session (see \hyperref[subsec:methods:controlenergy]{Methods}). 

Effective connectivity matrices were computed for each session and then averaged across sessions to obtain an overall effective connectivity matrix. For both monkeys, effective connectivity covariance increased across sessions (Monkey G: $f_{statistic}$ = 38.8, $p = 2.78 \times 10^{-7}$; Monkey Y: $f_{statistic}$ = 32.6, $p = 1.04 \times 10^{-7}$). A significant negative correlation between effective connectivity covariance and ACE was observed for Monkey G ($f_{statistic}$ = 16.2, $p = 2.65 \times 10^{-4}$ but not for Monkey Y \textbf{Supplementary Figure 6}). To mitigate the potential impact of channel variability across sessions, we chose to use the overall effective connectivity matrix rather than session-specific effective connectivity matrices for all control computations presented in the main text. \revtwo{We note that for MY, our ability to infer region-specific influences on network dynamics is limited by the variability in channel availability across sessions. For MG, all regions included in the analysis (CN, BA-8, and BA-9/45/46) were represented in all sessions. For MY, CN channels were available in 100.00\% of sessions, BA-8 channels were available in 97.27\% of sessions, BA24/25 channels were available in 17.27\% of sessions, and BA13/14 channels were available in 9.09\% of sessions.}

The ACE dynamics of both monkeys follow a similar downward trend throughout the entire experiment (\hyperref[fig4]{Figure 4C, 4D}). This finding is consistent with the notion that control energy reflects cognitive effort\cite{braun2021brain, zhou2023}; increasingly habitual behavior requires decreasing levels of cognitive control. A simple linear regression confirmed that there was a statistically significant negative relationship between ACE and session (Monkey G: $\beta$ = -0.1141 \revtwo{(95\% CI [-0.1703, -0.0578])},  $R^2$ = 0.30, $p = 1.98 \times 10^{-4}$; Monkey Y: $\beta$ = -0.0157 \revtwo{(95\% CI [-0.0235, -0.0080])}, $R^2$ = 0.13, $p = 1.06 \times 10^{-4}$). \revtwo{We note that the distributions of ACE across sessions differed significantly between the two monkeys, even after downsampling MY's dataset to match the size of MG's dataset (p = 0.0025926). Differences in data distribution could contribute to the differences in the strength of the relationship between ACE and session.} 

Notably, the drop in ACE across sessions as well as the correlations between ACE and other saccade metrics remained particularly robust across a range of different time horizons (0.5 to 1.5, in increments of 0.25; \textbf{Supplementary Figure 4}). \revtwo{These relationships were also robust to simulated increases in network size. We recomputed the energy cost associated with transitioning between each pair of brain states after simulating networks of 25, 100, 200, or 1,000 additional nodes for each monkey and embedded the original network in each of these simulated networks. The relationships between ACE and session and between ACE and saccade metrics persisted in the presence of additional nodes, suggesting that the results reported here hold when the circuit under study
is embedded in a wider network (Supplementary Figure 5).}

Though significant, we note that correlations between ACE and session were relatively low (\hyperref[fig4]{Figure 4C, 4D}). We believe this may be attributable to several factors. We expect that decreases in ACE underlie increases in habitual behavior, which does not increase in a perfectly linear fashion over time. Fluctuations in learning, and therefore in the control energy for neural state transitions, could be attributable to day-to-day fluctuations in motivational state. Daily fluctuations in motivation may be attributable to changes in internal state, including fluctuations in satiety\cite{berridge_reward_comp}.

\subsection*{Relating the Control Energy to Saccades}

\begin{figure*}[p] 
\centering
\includegraphics[width=0.8\textwidth]{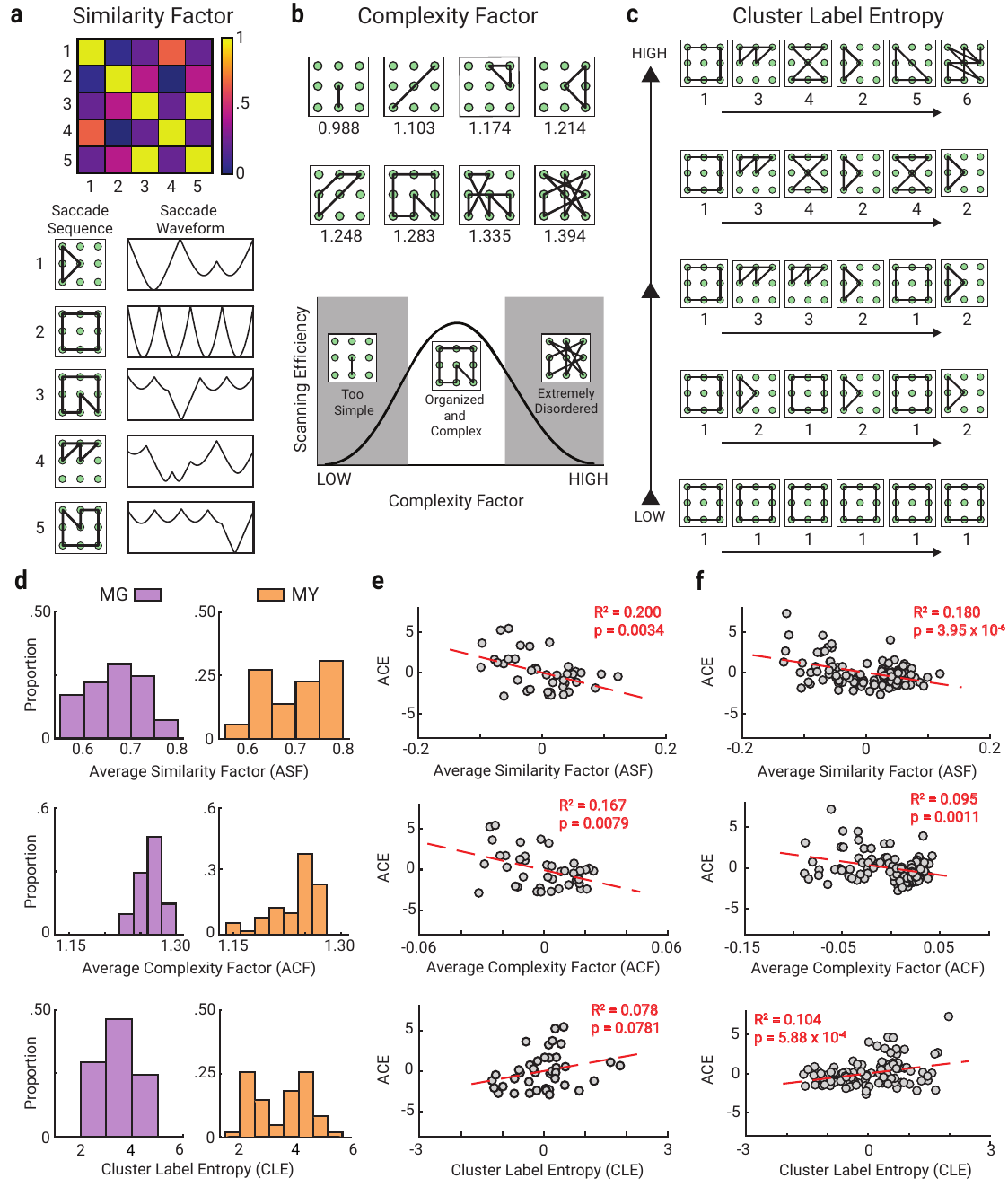}
\caption{\textbf{Relating Control Energy to Saccades}.  \textbf{(a)} The similarity factor (SF) captures information about the similarity between chronologically ordered trial saccade patterns. The similarity between two trial representative saccade patterns is calculated as a metric based on the Euclidean distance between two saccade waveforms. Here we show a visual depiction of 5 example patterns performed by the monkeys, their respective saccade waveform representations, and their similarity to one another. \textbf{(b)} The complexity factor (CF) is calculated as the fractal dimension of the saccade pattern. A range of observed patterns and their CFs are shown. Both extremely low and extremely high complexity patterns result in poor scanning efficiency during the task. \textbf{(c)} The cluster label entropy (CLE) metric captures information about the monkey's preference towards exploration of various patterns or exploitation of only a select few during a task session. The CLE is calculated as the Shannon entropy of the vector of identified trial cluster labels from a single session. Higher values indicate preference for constantly exploring a variety of clusters with minimal exploitation of any single cluster. \textbf{(d)} Side-by-side comparison of saccade metric distributions across all sessions for both monkeys. The average similarity and average CFs were calculated on a per-session basis. \textbf{(e-f)} Linear correlations between all average saccade metrics and the average minimum control energy (ACE) for Monkey G \textbf{(e)} and Monkey Y \textbf{(f)}, after adjusting for the number of channels in the recorded regions. The Coefficient of Determination ($R^{2}$) and $p$-value are provided for each plot. Each point represents one session.}
\label{fig5}
\end{figure*}

We next sought to quantitatively characterize how the monkeys' approaches to the free-scanning task changed over time. For this purpose, saccade patterns were characterized according to three metrics: the similarity factor (SF), complexity factor (CF), and cluster label entropy (CLE). These metrics were chosen because they reflect key properties of optimal habitual behavior: automatization and efficiency\cite{graybiel2015striatum}. More specifically, the SF reflects the repetitiveness of saccade patterns over trials, such that increased similarity may be understood to represent increased automatization as habitual behavior is acquired. The CF, or complexity of a saccade pattern, is inversely related to scan efficiency which increases with habit formation. Finally, the CLE reflects the degree of randomness with which the monkeys chose to perform particular saccade patterns. It is inversely related to scan repetitiveness which is associated with habitual behavior. Below we will discuss results for each metric in turn.

The first metric, the SF, represents the similarity between two TRSPs performed consecutively within the same session (see \hyperref[subsec:methods:classif]{Methods}). This metric can be used to answer the question, ``Is the monkey performing increasingly similar patterns the longer she engages in the task?''. The higher the value of this metric, the more similar the saccade patterns between trials. It is important to note that this metric was designed to be orientation-independent and reflection-independent. Accordingly, the SF renders two instances of the same pattern as identical even if one was rotated, the patterns were exact mirror images of each other, or rotations of exact mirror images (see \textbf{Supplementary Figure 7}). This feature of the similarity metric is shown in (\hyperref[fig5]{Figure 5a}), where patterns 3 and 5 only differ in their orientation but result in a SF of approximately 1. As hypothesized, the average SF persistently increased the longer the monkeys engaged with the task (Monkey G: $\beta$ = 0.0018 (95\% CI [0.0005, 0.0032]), $R^{2}$ = 0.17, $p = 0.0077$; Monkey Y: $\beta$ = 0.0013 (95\% CI [0.0011, 0.0014]), $R^{2}$ = 0.71, $p = 1.6 \times 10^{-30}$; \textbf{Supplementary Figure 8a}).

Although the range of similarity values across task sessions for both monkeys was the same (0.55 to 0.80), the ASF distribution of Monkey Y is skewed towards higher values (\hyperref[fig5]{Figure 5d}). A two-way Kolmogorov-Smirnov test confirmed that the ASF distributions of the two monkeys were significantly different from each other ($ks_{stat} = 0.3541$, $p = 7.57 \times 10^{-4}$). Furthermore, the Pearson correlation between the ASF and ACE (for Monkey G, $r = -0.447$, $p = 3.4 \times 10^{-3}$; for Monkey Y, $r = -0.424$, $p = 3.95 \times 10^{-6}$) was significantly negative in both monkeys, after adjusting for the number of channels in the recorded regions (\hyperref[fig5]{Figure 5e, 5f}). Permutation tests were performed independently for each monkey, to ensure that the observed associations between ASF and ACE were due to the observed neural circuit architecture (see \hyperref[subsec:methods:acecorr]{Methods}). The observed correlations between ASF and ACE for both monkeys proved to be significantly more negative than expected in their respective permutation null distributions (for Monkey G, $p < 10^{-3}$; for Monkey Y, $p < 10^{-3}$). On the basis of these results, we conclude that decreasing minimum control energy requirements are associated with the performance of increasingly similar saccade patterns.

In contrast to the SF, which captures repetitiveness, the second metric, the CF, quantifies the complexity of an individual TRSP within a session, defined as the fractal dimension of the pattern (see \hyperref[subsec:methods:classif]{Methods}). The metric can be used to answer the question, ``Is the monkey approaching the task in a strategic way or is it simply saccading at random?'' Both extremely low and extremely high complexity values are not optimal strategies for scanning the task grid efficiently. A pattern with a low complexity ($\approx$ 1) is often too simple and does not cover all the targets in the task. In contrast, a saccade pattern of high complexity (i.e., greater than 1.3) is extremely disordered, tortuous, and seemingly random without any strategy (\hyperref[fig5]{Figure 5B}). Patterns that strike a balance between organization and complexity offer the most efficient approach to scanning the 3$\times$3 target grid.

Both monkeys performed saccade patterns of similar complexity throughout their respective trials, with most sessions averaging to values between 1.24 and 1.28 (\hyperref[fig5]{Figure 5d}). However, the full range of complexity that Monkey Y exhibited was larger than that exhibited by Monkey G, as Monkey Y spent several sessions performing markedly simple patterns. Indeed, Monkey Y’s average CF (ACF) increased over time ($\beta$ = 0.0006 (95\% CI [0.0005, 0.0007]), $R^2$ = 0.55, $p = 1.79 \times 10^{-20}$; \textbf{Supplementary Figure 8b}). A two-way Kolmogorov-Smirnov test confirmed that the ACF distributions of the two monkeys were significantly different from each other ($ks_{stat} = 0.4581$, $p = 3.75 \times 10^{-6}$). Furthermore, the Pearson correlation between the ACF and minimum control energy (for Monkey G, $r = -0.409$, $p = 7.9 \times 10^{-3}$; for Monkey Y, $r = -0.307$, $p = 1.1 \times 10^{-3}$) was significantly negative in both monkeys, after adjusting for the number of channels (\hyperref[fig5]{Figure 5e, 5f}). Permutation tests were performed independently for each monkey, to ensure that the observed associations between the ACF and ACE were due to the observed neural circuit architecture (see \hyperref[subsec:methods:acecorr]{Methods}). The observed correlations between the ACF and ACE for both monkeys proved to be significantly more negative than expected in their respective permutation null distributions (for Monkey G, $p < 10^{-3}$; for Monkey Y, $p < 10^{-3}$).

Finally, the third metric, the CLE, is a quantitative estimate of a monkey's preference towards pattern exploration or exploitation during a task session. It is a direct calculation of Shannon's information entropy of the vector of chronologically ordered trial cluster labels in an individual session. The higher the CLE of a session, the less ordered the behavior and the more prone the monkey was to explore a variety of different saccade patterns coming from multiple identified clusters. The metric can be used to answer the question, ``Is the monkey choosing to explore many different saccade patterns across trials or does it continuously exploit a select few?'' Over time, both monkeys persistently decreased their CLE, exploiting a smaller number of saccadic patterns as the sessions went by (Monkey G: $\beta$ = -0.0491 (95\% CI [-0.0615, -0.0367]), $R^2$ = 0.62, $p = 8.78 \times 10^{-10}$; Monkey Y: $\beta$ = -0.0170 (95\% CI [-0.0198, -0.0141]), $R^2$ = 0.56, $p = 3.74 \times 10^{-21}$; \textbf{Supplementary Figure 8c}).

The distribution of CLE for Monkey Y across sessions shows that she exhibited moments of both extreme exploitation (CLE = 2-3) and extreme exploration (CLE = 4-5.5). In contrast, Monkey G exhibited a preference for mid-range values of CLE with a majority of the task sessions falling in the range of 3-3.5, a balance between exploitation and exploration (\hyperref[fig5]{Figure 5d}). A two-way Kolmogorov-Smirnov test confirmed that the CLE distributions of the two monkeys were not significantly different from each other ($ks_{stat} = 0.2020$, $p = 0.1539$).
The Pearson correlation between the CLE and minimum control energy (for Monkey G, $r = 0.278$, $p = 0.0781$; for Monkey Y, $r = 0.323$, $p = 5.88 \times 10^{-4}$) was significantly positive in Monkey Y only, after adjusting for the number of channels (\hyperref[fig5]{Figure 5e, 5f}). Permutation tests were performed independently for each monkey, to ensure that the observed associations between cluster entropy and ACE were due to the observed neural circuit architecture (see \hyperref[subsec:methods:acecorr]{Methods}). The significant correlation between CLE and ACE found in Monkey Y proved to be significantly more positive than expected from the respective permutation null distribution (for Monkey G, $p = 1$; for Monkey Y, $p < 10^{-3}$).

\revtwo{As noted above, Kolmogorov-Smirnov tests revealed significant differences in ASF and ACF (p < 0.001 for both), indicating that the monkeys exhibited distinct behavioral tendencies. Despite these differences, the relationship between ACE and each behavioral measure was consistent in direction across monkeys (Figure 5), suggesting a common underlying association. The observed distributional differences likely contribute to the variation in model fit statistics (e.g., $R^2$ values) between monkeys, and we highlight them here to aid interpretation of individual differences described throughout the manuscript.}

\revtwo{A hallmark of habit learning is a reduction in the cognitive load associated with performing the same behavior. To assess whether similar behavioral patterns become less costly over the course of learning in this task, we identified pairs of behavioral patterns with the same complexity factor that occurred in both early and late sessions for each monkey. For each matched pair, we obtained the control energy associated with the neural state transition between the two patterns for both early and late sessions. When a given transition occurred multiple times within an epoch (early or late), we averaged the control energy across all instances, yielding one value per transition type per epoch. This analysis revealed that the energy cost of transitioning between matched pairs was significantly lower in later sessions compared to earlier sessions for both monkeys (MG: $t(226) = 4.0612$, $p = 6.733 \times 10^{-5}$; MY: $t(824) = 6.657$, $p = 5.110 \times 10^{-11}$; \textbf{Supplementary Figure 11}), where the reported degrees of freedom correspond to the number of unique transition types observed in both early and late epochs. This finding is consistent with the interpretation that, over the course of training, transitions between similar behavioral states become more energetically efficient, reflecting reduced neural control demands. Such a reduction is a hallmark of habitual, automated behavior, and thus these results directly support our claim that habit learning in this task is accompanied by reduced cognitive load.}

Lastly, to ensure that there was no significant bias introduced by using the same sequence of spikes to infer both effective connectivity and control energy across sessions, we repeated the correlational analyses reported in this section by inferring effective connectivity on the first half of each session's trials and ACE on the remaining trials (\textbf{Supplementary Analysis 3}). Our results remained virtually identical. Taken together, all aforementioned results suggest individual differences \revtwo{in the distribution of exploitative and exploratory behaviors, yet a consistent direction of the relationship between these strategies and minimum control energy across monkeys, even though not all relationships reached statistical significance in both animals.}

\begin{figure*}[!tp]
\includegraphics[width=\textwidth]{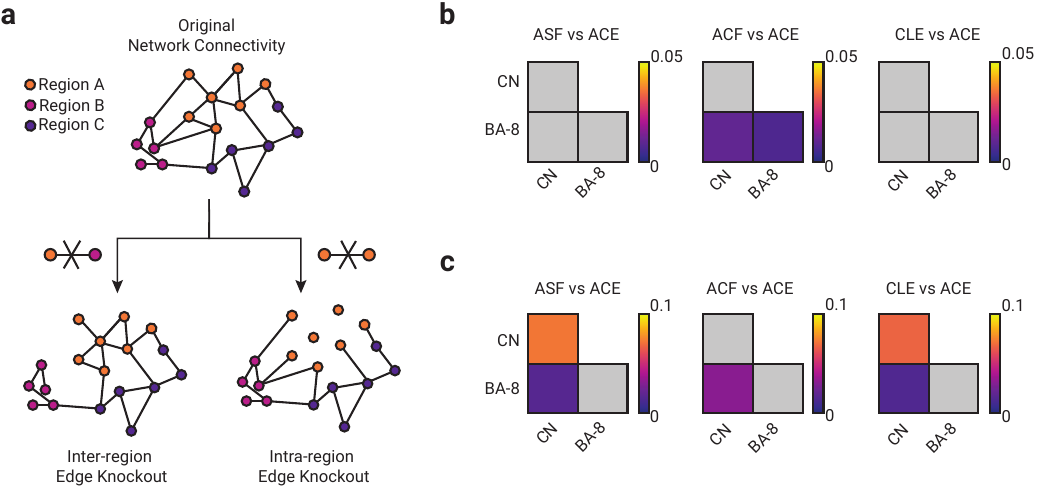}
\caption{\textbf{Virtual Region Specific Lesion Analysis.} \textbf{(a)} Visual depiction of lesion analysis workflow. The lesion knockout consists of performing a series of edge-knockouts where (i) all edges between two regions are set to zero in the effective connectivity matrix (inter-region), or where (ii) all edges connecting one region to itself are set to zero (intra-region). The average minimum control energy is then calculated using the lesioned effective connectivity matrix, and correlations are computed between the average minimum control energy and all saccade metrics. \textbf{(b-c)} Results of lesion analysis across \revtwo{regions of the CN and BA-8} for Monkey G \textbf{(b)} and Monkey Y \textbf{(c)}. All post-lesion correlations were compared to an equivalent null distribution constructed from random edge lesions. Matrix elements represent the absolute value difference between post-lesion and pre-lesion correlations. All grayed out edges represent lesions which did not result in a significant change in correlation value when compared to their respective null distribution. ASF: Average Similarity Factor; ACF: Average Complexity Factor; CLE: Cluster Label Entropy; ACE: Average Control Energy; CN: Caudate Nucleus; BA: Brodmann Area}
\label{fig6}
\end{figure*}

\subsection*{Simulation Analysis}

Given that neurons found in the brain areas examined in this work have been found to display directional tuning\cite{Crapse2009frontal,messinger2021frontal}, and provided that neurons’ directional tuning can affect their firing rates depending on the directionality of saccadic movements in space (as in the case of a grid), we performed a simulation to ensure that our main results reported in the previous sections are not driven by directional tuning alone. This was particularly important in our study since the neurons’ firing rates were used to generate the effective connectivity matrices that were then used to compute the control energy metrics. In brief---and as explained in further detail in \hyperref[subsec:methods:acecorr]{Methods}---we generated ``null'' effective connectivity matrices for each Monkey using only the Monkey’s saccadic movements and not her measured firing rates; these matrices were then used to compute each Monkey’s ``simulated'' average minimum control energy. We first examined how each Monkey’s simulated average control energy changed over sessions. Even though there was a significant negative correlation between the empirically-measured ACE and session progression, the relationship between the simulated ACE and session progression was not significant (Monkey $G$: $\beta$ = -0.061 (95\% CI [-0.136 0.015]), $R^2$ = 0.063, $p$ = 0.1122; Monkey $Y$: $\beta$ = -0.005 (95\% CI [-0.015 0.006]), $R^2$ = 0.008, $p$ = 0.3613). 

We next compared the simulated ACE with the empirically-derived ASF. There was once again a lack of association between the two variables for both Monkeys (for Monkey $G$, $r$ = -0.233, $p$ = 0.1430; for Monkey $Y$, $r$ = -0.179, $p$ = 0.0607), indicating that directional tuning alone could not explain the significant association reported between the empirical variables. Repeating these analysis between the simulated ACE and ACF, we found that there was no association between the two variables in one of the two monkeys (Monkey $G$, $r$ = -0.280, $p$ = 0.0767). Even though the association between ACF and the simulated ACE remained significant for the other Monkey (Monkey $Y$, $r$ = -0.275, $p$ = 0.0037), the effect size was notably lower than that between the empirical variables. Lastly, we compared the simulated ACE with the CLE and found no significant correlation between the two variables in both Monkeys (for Monkey $G$, $r$ = 0.159, $p$ = 0.3222; for Monkey $Y$, $r$ = 0.115, $p$ = 0.2317). Collectively, these results lend support to the notion that directional tuning alone could not explain the significant associations reported between the empirical variables.

\subsection*{Identifying Neural Substrates Particularly Key to the Relation Between Control Energy and Saccades}

In a final step, we seek to determine which part(s) of the inferred network of brain regions significantly contribute to the observed relationships between control energy and behavior. We do so by performing a virtual lesion analysis consisting of a series of inter- or intra-region edge knockouts in the inferred effective connectivity matrix (see \hyperref[fig6]{Figure 6A} and \textbf{Supplementary Figure 9} for a schematic depiction of this approach). An edge knockout refers to setting the weights of edges in the effective connectivity matrix to a value of zero, thereby virtually removing connections in the network. Edges whose knockout serves to remove the correlation (resulting in a $p$-value greater than $\alpha=0.05$) between ACE and saccade metrics are inferred to be important in controlling task specific energy dynamics. If the correlations can be removed by localized edge deletions, then we would infer that the energetic constraints on neural state transitions are localized to a particular part of the circuit. If instead the correlations cannot be removed by localized edge deletions, then we would infer that the energetic constraints on neural state transitions are broadly distributed across the circuit.

We first performed inter-region and intra-region edge knockouts across all possible combinations of brain regions for each monkey. Inter-region edge knockouts consisted of setting all edges between two regions to zero in the effective connectivity matrix, while intra-region edge knockouts consisted of setting all edges connecting one region to itself to zero. The results are shown in \hyperref[fig6]{Figure 6B} and \hyperref[fig6]{Figure 6C}. In Monkey G, the removal of connections between BA-8 and BA-9/45/46 resulted in small but significant changes to the originally observed correlation between the control energy and the ASF as well as the ACF. In Monkey Y, the removal of connections within the caudate nucleus resulted in small but significant changes to the originally observed correlation between the control energy and both the ASF and CLE. Although significant changes were found in both monkeys, none were drastic enough to fully disrupt the observed correlations between behavior and control energy. 

In order to directly assess the relative importance of inter- versus intra-region knockout, we performed 1,000 virtual knockouts of 500 random inter-region edges and 1,000 virtual knockouts of 500 random intra-region edges, computed the change in correlation between ACE and behavior for each knockout, and compared the two resulting distributions for each monkey. For Monkey G, eliminating inter-region edges had a significantly greater impact on the correlation between ACE and ACF than removing intra-region edges ($d$ = 1.4348, $p = 0.0030$), while there were no differences in the impact of inter- versus intra-region edge lesions on the correlations between ACE and ASF or CLE. For Monkey Y, there were no differences in the impact of inter- versus intra-region edges lesions on correlations between ACE and any of the saccade behavioral metrics. In general, we note that differences in the impact of inter- versus intra-region edge lesions on the relationship between control energy and behavior are limited and individual-specific.

The observed behavior-energy correlations exhibited resilience to disruptions in the inferred effective connectivity (See \textbf{Supplementary Figure 10}). In order to successfully disrupt the observed correlations in Monkey G, a minimum of 84\% of all edges from its connectivity matrix have to be removed. This minimum threshold increases for Monkey Y, where at least 95\% of all edges have to be set to zero in order to significantly disrupt the correlations. These findings suggest that the energetic constraints on neural state transitions relevant for behavior are only partially localized (\hyperref[fig6]{Figure 6B} \& \hyperref[fig6]{Figure 6C}), but may be more accurately described as being broadly distributed across the circuit.

\section*{Discussion}

Learning commonly requires the development of strategies to increase reward in the face of uncertainty\cite{gottlieb2018towards}. Such strategies can be manifested in sequential behaviors that serve to continuously gather information about the environment\cite{desrochers2010optimal}. Yet precisely what rules guide the formation of sequential behaviors remains poorly understood. Although recent work highlights the relevance of distributed cortico-striatal circuitry\cite{smith2013dual}, progress has been hampered by the lack of a formal theory linking activity in such circuitry to habitual (or non-habitual) behavior. Here we address this challenge by employing network control theory to determine the energy requirements of sequences of neural states occurring atop a complex network structure. 

Combining behavioral measurements and neural recordings from two female macaque monkeys performing a free-view scanning task over 60–180 sessions\cite{desrochers2015habit,desrochers2010optimal}, we find that smaller energy requirements are associated with transitions between trials in sessions with a high degree of similarity between complex saccade patterns, and in sessions characterized by an emphasis on the repetition of a small subset of patterns rather than exploration of a more diverse set of distinct patterns. \revtwo{We also find that monkeys tend to rely on scan patterns of intermediate complexity, and that within this limited range, complexity and control energy are inversely related. Together, these findings are consistent with the principle of maximum entropy, as they suggest that the distribution of scan patterns maximizes unpredictability over the feasible set of states defined by energetic and behavioral constraints, favoring low-energy, intermediate-complexity patterns while avoiding the extremes of very simple or very complex patterns.} Moreover, we employ a virtual lesioning approach to demonstrate that the derived relationships between control energy and behavior are highly resilient to small, local disruptions in the network, suggesting these observations are associated with the network as a whole rather than a small subset of its nodes. 

Earlier studies of these data provide evidence for learning of habitual behavior in the absence of instruction and define neural signals underlying the development of habitual behavior sequences. Saccade patterns were found to become increasingly repetitive and lower in entropy over the course of repeated task trials\cite{desrochers2010optimal}, and neurons within the caudate nucleus were found to signal both cost and outcome in the context of habit learning\cite{desrochers2015habit}. Our work builds on these previous studies by determining how interactions between regions of the frontal cortex and caudate nucleus give rise to habit learning and highlights individual differences in the spatially distributed nature of neural activity that drives learning. More broadly, our study advances a theoretically principled approach to the study of habit formation, provides empirical support for those theoretical principles, and offers a blueprint for future studies seeking to explain how behavior arises from changing patterns of activity in distributed neural circuitry. 

The study of habit learning, like the study of many other cognitive functions, has benefited immensely from lesion studies\cite{teng2000contrasting,price2003lesion,izquierdo2004lesion,jenrette2019lesions} and from focused recordings in single brain regions\cite{yassin2010neocortex,yanike2014representation,kim2015dopamine,desrochers2015habit}. Yet, the field has long appreciated that single regions do not act in isolation, but instead form key components of wider circuits relevant for perception\cite{whitmire2016rapid,garvert2014amygdala}, action\cite{makino2016circuit}, and reward\cite{cox2019striatal}, among others. With recent concerted funding support\cite{litvina2019brain}, many new technologies are now available for large-scale recording of neural ensembles, including methods for high-density multi-region recordings\cite{feingold2012system,chung2019high} and associated novel electrode technologies\cite{hong2019novel}. These advances support a wider goal to gather evermore detailed measurements of activity across the whole brain\cite{kleinfeld2019can}. Here, we capitalize upon such technological advances, and utilize multi-channel, multi-area recordings to better understand the distributed nature of neural circuitry underlying habit formation. The channels span Brodmann areas 8 (frontal eye fields), 9 (dorsolateral and medial prefrontal cortex), 13-14 (insula and ventromedial prefrontal cortex), 24-25 (anterior and subgenual cingulate), and 45-46 (\emph{pars triangularis} and middle frontal area), allowing us to probe multi-area activity and inter-areal interactions that track habitual behavior.

Our data naturally motivate the question of how circuit activity supports behavior. This question is certainly not new, and not even specific to neural systems. In fact, the recent rapid expansion of work in artificial neural networks has highlighted the fundamental fact that the architecture of a network is germane to the system's function\cite{veen2019neural}. Liquid state machines\cite{maass2002real}, convolutional neural networks\cite{lecun1998gradient}, and Boltzmann machines\cite{hinton1986learning} all perform distinct tasks defined by their architectures. Similarly, in biological neural systems, empirical and computational evidence links the architecture of projections with the nature of memory retrieval\cite{rajesthupathy2015projections}, flexible memory encoding\cite{curto2012flexible}, sequence learning\cite{rajan2016recurrent}, and visuomotor transformation\cite{murdison2015computations}. Intuition can be drawn from simple small architectures or network motifs\cite{kashtan2005spontaneous} which have markedly distinct computational and control properties\cite{whalen2015observability}. For example, a chain is conducive to sequential processing, whereas a grid is more conducive to parallel processing. Unfortunately, the architecture of multi-area circuits in the primate brain is not quite so simple, thus hampering basic intuitions and straightforward predictions. To address this challenge, we use the mathematical language of network science\cite{mitchell2011complexity}. The network approach allows us to embrace the distributed nature of neural circuit activity and quantitatively describe the empirically observed architecture, while also formalizing questions regarding how that architecture supports circuit function\cite{bassett2018on}.

Current efforts in computational and systems neuroscience are divided by a focus either on patterns of neural activity or on patterns of neural connectivity. At the small scale, this divide separates studies of the firing rates of neurons from studies of noise correlations\cite{doiron2016mechanics,kohn2016correlations}. At the large scale, this divide separates studies using general linear models or multivoxel pattern analysis in fMRI\cite{friston2005models,mahmoudi2012multivoxel} from studies using graph theoretical or network approaches\cite{bassett2018on}. A key challenge facing the field is the need to span this divide, both in experimental and in theoretical investigations\cite{ocker2017linking}. Indeed, to take the next step in understanding behavior requires the development of computational models of cognitive processes that conceptually or mathematically combine activity and connectivity\cite{bassett2017network}. 

Here we summarize firing rate activity across channels as a brain state, and we probe how such states can change given the effective connectivity between channels. The fact that network architecture can constrain the manner in which activation patterns change (and \emph{vice versa}\cite{schuecker2017fundamental}) is supported by empirical evidence in large-scale human imaging\cite{ito2019discovering}, and a long history of computational modeling studies in human and non-human species\cite{breakspear2017dynamic,ocker2017linking}. Here we inform our modeling choice by noting a particular characteristic of that constraint, which exists in the following form: state $x$ is more likely than state $y$ given network $A$. Specifically, we acknowledge that neural units that are densely connected are more likely to share the same activity profile than neural units that are sparsely connected, for example due to being located in a distant area. This phenomenon naturally arises in many dynamical systems\cite{fan2019enhancing,sorrentino2016complete,TM-GB-DSB-FP:18}, and its study has recently been further formalized in the emerging field of graph signal processing\cite{ortega2018graph,huang2018graph}, which offers quantitative measures to evaluate the statistical relations between a pattern of activity and an underlying graph. 

Beyond positing a probabilistic relationship between activity and connectivity, we define an \emph{energetic} relation between them. Our formalization uses the nascent field of network control theory\cite{motter2015network,pasquletti2014controllability,liu2011controllability}, which develops associated theory, statistical metrics, and computational models for the control of networked systems, and then seeks to validate them empirically. The broad utility of the approach is nicely exemplified in recent efforts that address such disparate questions as how to control the spread of infectious disease in sub-Saharan Africa\cite{roy2011network}, of current in power grids\cite{barany2004nonlinear}, or of pathology in Alzheimer's disease\cite{henderson2019spread}. In the context of neural systems, the theory requires three components: (i) a definition of the system's state, such as population-level activity in large-scale cortical areas\cite{stiso2019white} or cellular activity in microscale circuits\cite{yan2017network}, (ii) a measurement of the connections between system components, such as white matter tracts\cite{bernhardt2019temporal} or synapses\cite{towlson2018caenorhabditis}, and (iii) a form for the dynamics of state changes given a network, such as full non-linear forms\cite{schiff2010neural} or linearization around the current operating point\cite{jeganathan2018fronto}. Here we let the state reflect the firing rate activity across channels, the network reflect effective connectivity between channels, and the dynamics take the form of a linearization around the current operating point. 

After formalizing the theory for a given system, one can use longstanding analytical results to estimate the energy required to move the system from one state to another\cite{hespanha2018linear,boltyanskii1960theory}. Our approach assumes that the system is linear. Though this assumption is inconsistent with neural dynamics at the microscale cellular level, which are non-linear, prior work indicates that linear models more effectively approximate macroscale neural dynamics than non-linear models and are more readily interpretable\cite{nozari2020brain}. A linear model is therefore suitable for modeling the population-level neural dynamics represented in our data. 

We focus specifically on the problem of identifying the minimum control energy, which is a common subform of the more general problem of identifying the control energy required in the optimal trajectory between state $i$ and state $j$\cite{kailath1980linear}. Outside of neuroscience, the approach has proven useful, for example, in increasing energy efficiency in induction machines\cite{plathottam2015transient}, enhancing performance of transient manufacturing processes\cite{sahlodin2017efficient}, and managing energy usage in electric vehicles\cite{boehme2013trip}, among others. In the context of neural systems, the study of such trajectories has been used to address questions of how the brain moves from its resting or spontaneous state to states of task-relevant or evoked activity, how the brain's network architecture determines which sets of states require little energy to reach, and how electrical stimulation induces changes in brain state\cite{stiso2019white}. Here, we use the approach to estimate the amount of energy that is theoretically needed to push the circuit from the state reflecting firing rate activity in one trial to the state reflecting firing rate activity in the next trial. By performing the calculation for all pairs of temporally adjacent trials, we are able to examine changes in energy over the course of learning. More importantly, structuring the investigation in this way allows us to determine how such energy relates to pairwise differences in sequential behaviors during habit formation.

In an expansion upon prior work in the application of network control theory to neural systems, we posit that low energy state transitions characterize processes (and their associated behaviors) that are less cognitively demanding. Informally, the underlying notion harks back across at least two centuries in the history of neuroscience\cite{sourkes2006on}. Formally, we can draw insight from the principle of maximum entropy\cite{ortega2013thermodynamics}: when estimating probability distributions (or transitions in this case), the probabilistic transitions that leave the largest remaining uncertainty in the system (i.e., the maximum entropy) are naturally the most unbiased ones in representing the system. Based on this principle, we can thus posit that transitions between low entropy saccade patterns require greater effort and therefore energy than transitions between high entropy saccade patterns. Note that a high entropy saccade pattern is one that spans many targets in a disordered manner while a low entropy saccade pattern is one that spans few targets in an organized, structured pattern. 

Concretely, we operationalize the entropy of a TRSP (not to be confused with the monkeys’ cluster label entropy that quantifies the extent to which the monkey selects saccade patterns from trial to trial in an ordered fashion) as its fractal dimension, which we refer to as its complexity. Consistent with our hypothesis, we find that the saccade complexity is negatively correlated with the energy theoretically required to move the neural circuit from the firing rate state of one trial to the firing rate state of the next trial. Broadly, our data join those acquired in other model systems, anatomical locations, and species in providing evidence that maximum entropy models explain key features of neural dynamics\cite{savin2017maximum,granot2013stimulus,meshulam2017collective}. 

Moving beyond the assessment of a saccade pattern's entropy, we next consider the role of habits in modulating the cognitive demands elicited by a task\cite{haith2018cognitiveload,moors2006automaticity}. We hypothesize that transitioning between the same (or similar) saccade patterns will require less cognitive effort, and therefore less energy, than transitioning between different (or dissimilar) saccade patterns. Consistent with our hypothesis, we find that transitioning between more similar saccades is associated with smaller predicted energy, and that sessions with a larger number of distinct saccade patterns are associated with greater predicted energy. Collectively, these data comprise a formal link between the energetics of neural circuit transitions and sequential behaviors.

By definition, the theoretically predicted energy is a function of both the neural states and the underlying network of effective connectivity, and therefore reflects contributions from all channels and from all inter-channel relations. Nevertheless, it is still of interest to determine whether some channels, or some inter-channel relations, contribute relatively more or less than others, as such regions (or edges) can be targeted with external stimulation to alter behavioral output and potentially guide clinical practice\cite{stiso2019white}. Using a virtual lesioning approach, we found that removal of connections in the caudate nucleus, and connections between BA-8, BA-9/45/46 and BA-13/14, resulted in predicted energies that caused small but significant decreases in magnitude to the correlation with saccade metrics. 

Our work builds on prior studies by highlighting the importance of network-level coordination of cortical and subcortical regions for habit learning behavior. The caudate nucleus, frontal eye fields, and prefrontal cortex have been shown to play distinct roles in learning and saccade behavior. Specifically, the frontal eye fields and other regions of the cortex are involved in saccade target selection\cite{iwamoto2010}. The role of prefrontal cortex is consistent with transcranial magnetic stimulation studies showing its necessity for higher-level sequential behavior\cite{desrochers2015necessity} and its involvement in uncertainty driven exploration\cite{badre2012rostrolateral}. 

The critical role of the caudate nucleus in habit formation is consistent with prior lesion studies\cite{teng2000contrasting,price2003lesion,izquierdo2004lesion,jenrette2019lesions} and recording studies\cite{yassin2010neocortex,yanike2014representation,kim2015dopamine,desrochers2015habit}. Neurons within the caudate nucleus have previously been shown\revtwo{--using the same dataset analyzed in the present study--}to encode both the cost and benefit of performing a given sequence of behaviors\cite{desrochers2015habit} and their activity is correlated with successful learning of stimulus-outcome associations\cite{seger2005CN}. Among the edge type lesions that were found to significantly impact the relationship between control energy and saccade metrics, most (50\% for Monkey G and >85\% for Monkey Y) involved the caudate nucleus. This finding furthers our understanding of the critical role of the caudate nucleus in learning stimulus-outcome associations and informing the development of habitual behavior. 

The caudate nucleus and cortex interact via cortico-striatal loops\cite{seger2009loops}, and the caudate nucleus is thus anatomically positioned to receive input from as well as to provide feedback to cortical regions that guide saccade target selection. Our findings build on those of prior studies by demonstrating that virtual lesioning of connections both within and between regions of the cortex and caudate nucleus reduces correlations of ACE with saccade behavioral metrics. Our subsequent random lesioning results further extend our understanding of the neuroanatomical support for these behaviors by suggesting that the energetic constraints on neural state transitions are broadly distributed across the circuit. Furthermore, given that the majority of network edges needed to be lesioned for the behavior-energy correlations to be impacted, it could also be argued that the network architecture itself might not play as significant of a role in energetically constraining the neural state transitions. Future work should investigate this possibility, and examine whether the spatial scale (i.e., number of nodes) or architecture of the network determines how resilient its behavior-energy relationship is to external perturbations.

Lastly, another feature of our findings that we find perhaps particularly striking is their specificity to the two monkeys. Monkey G performed more dissimilar saccade patterns from trial to trial, consistent with the goal-directed exploration supported by prefrontal connections identified in our lesioning analysis. In contrast, Monkey Y performed more similar saccade patterns from trial to trial during sessions, consistent with lower-level habit formation supported by caudate nucleus identified in our lesioning analysis. While our study is underpowered to formally probe individual differences, it would be particularly interesting to examine how the energy-behavior relations described in this work differ across a larger number of individuals. A fruitful future endeavor would also include investigating how the energy-behavior relationships differ between healthy and diseased individuals.

Several methodological considerations are particularly pertinent to this work, and here we mention the three that are most salient. First, we note that in understanding the manner in which neural units communicate, one might wish to have full knowledge of the structural wiring between those neural units\cite{schafer2018worm,eichler2017complete}. Despite recent advances in technology at the cellular scale\cite{hillman2019light,eberle2015mission,berger2018vast}, such information is challenging to acquire \emph{in vivo} in large animals, and currently not possible at all in primates. A reasonable alternative is to use the empirical measurements of activity to \emph{infer} the effective relationships between neural units\cite{friston2011functional,tavoni2016neural,schiefer2018from}. Here we take precisely this tack, thereby distilling a weighted effective connectivity matrix summarizing the degree to which each channel statistically affects another.

It has been found that effective connectivity is generally shaped by anatomical connectivity\cite{sokolov2019linking,crimi2021structurally,friston2011functional}, and may therefore convey similar information about the network's structure. The use of effective connectivity in our model is also consistent with the notion that both structural connections and functional integration within neural systems contribute to their controllability\cite{scheid2020pnas}. A marked benefit of effective over structural connectivity between large-scale brain areas is that only the former can be used to study temporal variation on the time scale at which learning occurs\cite{battaglia2012dynamic}. Nevertheless, there are limitations associated with this approach. Changes in effective connectivity may occur over shorter time scales than changes in structural connectivity, particularly during learning, and the average effective connectivity matrix used to compute minimum control energy in this study may not completely capture the relations between brain regions. The neurons recorded may also change over time due to shifts in electrode positioning during the study, potentially leading to changes in effective connectivity that our approach does not capture. 

Finally, of relevance to this methodological consideration, missing or ``hidden'' nodes could impact the integrity of either structural or effective connectivity measures. \revtwo{Although the results of our sub-graph sampling simulation analysis suggest that increasing the number of nodes in the network does not disrupt the relationship between control energy and behavioral metrics (Supplementary Figure 5),} future work including recordings from a larger number of brain regions could \revtwo{experimentally validate the persistence of} the energy-behavior relationships studied in this work. Such an examination would be of particular interest in assessing whether the relationship between control energy and saccadic movements depends on the spatial scale of the network (i.e., number of cortical and subcortical brain regions observed). From a graph theoretical standpoint, such work could be used to inform whether properties of a subgraph sample can represent properties of the full graph from which they have been sampled.

A second important consideration pertinent to our work is that we use analytical results from the study of linear systems\cite{kailath1980linear} to inform our network control theory approach\cite{pasquletti2014controllability,stiso2019white}. It is well known that neural dynamics -- measured in distinct species and across several imaging modalities -- are in fact non-linear\cite{breakspear2017dynamic}. Linear models of non-linear systems are most useful in predicting behavior in the vicinity of the system's current operating point\cite{kim2019linear}, or for explaining coarse time-scale population-average activity\cite{honey2009predicting}. 

Of particular relevance to our results is the recent validation of a tight relationship between control energy and metabolic energy\cite{he2021pathological}, where increased minimum control energy was found to be correlated with hypometabolism as measured by FDG-PET. Of equal relevance to our results is the recent validation of the use of linear models for neural data\cite{nozari2020brain}, where linear models outperformed a wide breadth of non-linear models in predicting fMRI and iEEG data with minimal model complexity. Although a direct assessment of the performance of linear and nonlinear models for approximating neural dynamics measured by multi-unit electrode recordings has not has not yet occurred, we would expect the results of Nozari et al. to generalize due to the relevance of the four factors that were found to mask nonlinearities in that study: averaging over space and time, observation noise, and limited data samples. For the study of other sorts of behavior or signals, future work could consider extending our simulations to include appropriate nonlinearities\cite{motter2015network}. 

A third important consideration relevant to our work is the fact that animal behaviors in general -- and saccades in particular -- are complex and difficult to describe cleanly\cite{leshner2011quantification,bellet2019human}. Here we address this difficulty by developing a novel algorithmic approach to the extraction of representative saccade patterns. We specifically chose to represent repetitive, cyclic saccadic movements as 2-D polygons (i.e., TRSPs) drawn on the 3$\times$3 grid of equally sized and spaced dots presented during the task; every saccade was represented as a straight line between two dots on the grid. Given the setup of the task, this previously used approach\cite{desrochers2010optimal, desrochers2015habit} is able to faithfully capture a variety of different saccadic patterns visited. Broadly, our method capitalizes on a graphical representation of saccades, which in turn allows us to use previously developed tools for the characterization of graphs\cite{mitchell2011complexity}. 

There are limitations associated with this approach that warrant mentioning. First, this approach is based on the assumption that meaningful differences in behavior reflective of habit learning are associated with changes from one trial to the next; however, changes in behavior over the course of a trial are not represented. Second, saccade metrics are designed to be rotationally invariant; however, it is known that monkeys possess innate preferences for saccades in one direction. In future work, it would be of interest to assess whether the energetic cost associated with performing a saccade in a particular direction reflects this side bias—in other words, is it less costly to perform a saccade in the preferred versus the non-preferred direction? 

Finally, saccade complexity could be influenced by the arrangement of targets in the task, and this possible influence could limit the generalizability of the results. For instance, if targets were arranged in a circle instead of a grid, an ordered pattern of saccades that is low in complexity might be more effective than a complex pattern. If targets were instead arranged at random, a more complex pattern might be more energetically favorable. Of course, other representations—and further analyses—of saccadic trajectories can be employed, such as embedding the saccadic waveforms into higher-dimensional spaces. Our effort follows a growing literature using network models to parsimoniously represent and study animal behavior\cite{belyi2017global,hawkins2019emergence}. In the future, eye-tracking task setups capturing saccadic movements across a larger number of data points across the grid could be employed.

Systematically canvassing uncertain environments for reward induces habitual behaviors and engages distributed neural circuits. Here we offer a formal theory based on the principles of network control to account for how pairwise differences in sequential behaviors during habit formation can be explained by the energetic requirements of the accompanying neural state transitions. In doing so, the study frames the concept of cognitive computations within a formal theory of network energetics. Our findings further support the notion that the principle of maximum entropy could be a useful explanatory principle of behavior. While outside the scope of this study, many relevant questions remain unasked. Future work could usefully expand upon our observations by increasing the number of recorded regions or altering the task to include different sorts of environmental uncertainties. Incorporation of additional computational capabilities into the theory, and exercising agent-based simulations to determine optimal cost functions and associated learning rules for artificial neural systems placed in similar environments could also be used to expand upon this work.

\section*{Methods}
\label{sec:methods}

The data consist of behavioral measurements and neural recordings from two adult female monkeys (Maccaca mulatta, ~5.9kg each): Monkey G and Monkey Y. All procedures were performed as approved by the Massachusetts Institute of Technology’s Committee on Animal Care. Full descriptions are provided in earlier publications on this dataset\cite{desrochers2010optimal,desrochers2015habit}, and here we briefly summarize. 

Both monkeys were individually monitored, and data was recorded while the monkeys performed a free-viewing scan task. The task was performed across multiple days (sessions) with each session consisting of multiple back-to-back trials. Eye movements were recorded utilizing an infrared tracking system and converted into a sequence of saccades, or rapid eye movements from one point to another. All measurements of neural activity were obtained from individual chronically implanted electrode arrays recording bilaterally from various points in the caudate nucleus, frontal eye fields, and prefrontal cortex. \par

\subsection*{Recording Technology}

Prior to neural recordings, a head fixation post and recording chamber were placed in separate surgeries. All surgical procedures were performed under sterile conditions on deeply anesthetized monkeys placed in a standard stereotaxic apparatus. Each recording chamber was centered mediolaterally on the midline of the brain (with 20 mm on either side) and at approximately 22 mm (monkey G) or 30 mm (monkey Y) anterior to the interaural line.

The plastic recording chamber accommodated a 30 × 40 mm grid with holes spaced at 1 mm center to center. These holes accommodated custom-built screw microdrives (Specialty Machining) that could each carry three, six, or nine independently movable electrodes (1–2 M$\Omega$ at 1 kHz, 110-130 mm long, 125 $\mu$m shank, ~3 $\mu$m diameter tip; Frederick Haer). The microdrives could be configured to space electrodes as close as 1 mm apart and to record simultaneously from many brain regions, and their placements could be reconfigured for each chronic implant. Before the implantation of the chronic electrodes, electrophysiological mapping by the use of microstimulation (trains of 24-64 250 $\mu$s wide biphasic pulses, 333 Hz, 10-150 $\mu$A) was completed to determine the location of several key landmarks such as the FEF, primary motor cortex, and the depth of the dorsal surface of the CN.

Data from monkey G is from one chronic implant containing 72 independently movable electrodes targeting the CN, FEF, and PFC bilaterally. It was in place for 140 days. Both of the chronic implants (166 and 366 days, respectively) in monkey Y contained 96 electrodes and also bilaterally targeted the CN, FEF, and PFC.

The recording of all data (behavioral and neural) began on the first day of training after electrodes had been implanted. Eye position was monitored by an infrared eye tracking system (500 Hz; SR Research Ltd.). Custom behavioral control software was designed in Delphi III. All signals were recorded with a Cheetah Data Acquisition system (Neuralynx, Inc.). Spike waveforms were digitized at 32 kHz (filtered 600-6,000 Hz). Local field potentials were filtered at 1-475 Hz, and eye movements were recorded at 2 kHz.
    
\subsection*{Task Structure} 

The task begins when a grid of small gray circles is presented on a screen in front of the monkey, whose head is fixed in place. After a variable period of time, the inner gray circles of the grid are replaced with a 2$\times$2 grid or a 3$\times$3 grid of larger green dots (\emph{Targets On}). The monkey’s gaze may not leave the space defined by the perimeter of the green dots or the trial will be marked as unsuccessful and the screen will revert back to a grid of only gray dots. After a variable time, one of the green targets is baited (\emph{Target Baited}) such that if the monkey’s gaze falls on to the baited target, the trial is rewarded. The monkeys were not given information about when the target was baited or which target was baited. At this point in the task, if the monkey’s gaze crossed into or over the bait target, the grid of green dots was replaced by the original gray dots (\emph{Targets Off}). After a variable time, the monkey was presented with a short reward to indicate success; acknowledging their preferences, juice was offered to Monkey G, and an alternative reward mixture was offered to Monkey Y. \par

\subsection*{Data Quality Assurance and Cleaning}

Due to the inherent complexity of the task and behavioral responses thereto, it is critical to apply data quality standards that ensure statistical analyses are appropriate and well-powered. Accordingly, all analysis involved in inferring effective connectivity was limited to task trials where the monkeys were presented with the 3$\times$3 grid version of the free-scanning task. \revtwo{Early sessions in which the 3$\times$3 grid were not present were excluded from the analyses.} Furthermore, the number of available channels, defined as those with non-zero signal, varied across sessions. The 60 recorded sessions for Monkey G contained anywhere from 16 to 38 channels (with an average of 23) from a total of 72 unique channels. The 180 recorded sessions for Monkey Y contained anywhere from 3 to 23 channels (with an average of 11 channels) from a total of 96 unique channels. 

To ensure adequate sampling, all analysis performed on Monkey Y was limited to sessions containing 8 or more channels. These criteria resulted in 18,298 available trials for Monkey G and 157,729 for Monkey Y. Analytical steps focused on defining the relationship between task behavior and control energy dynamics were limited to only 3$\times$3 grid task trials that were rewarded and exhibited a looping saccade sequence, which is defined as a sequence that starts and ends at the same grid target. A total of 9,702 ($\sim$53\%) task trials were used for Monkey G and 80,664 ($\sim$51\%) for Monkey Y. In this particular task, a looping saccade sequence in which all targets were visited once before returning to the starting node is considered optimal\cite{desrochers2010optimal}.  \par

\subsection*{Classification of Saccade Sequences} \label{subsec:methods:classif}

Measurements of monkey eye-movements during the free-scanning task comprise a list of saccades performed in the scanning window of a given trial. Each saccade is represented as a vector of two numbers, which denote the start and end grid targets of the saccade. The entire sequence of saccades performed during a given trial can be written as an $m \times 2$ matrix, $SL$. Each row in $SL$ is a single saccade, with the first and second column representing the start and end targets, respectively. This representation can be thought of as a list of directed connections between points. It is then possible to convert a trial specific sequence of saccades into a directed and weighted graph, which we will refer to as the \emph{saccade network}. 

In graph theory\cite{mitchell2011complexity}, a generic network is made up of $N$ nodes that are connected pairwise by $E$ edges. Here, a saccade network consists of nine nodes ($N = 9$), one for each grid target, and edges defined by the saccade sequence. Specifically, an edge between two nodes in a saccade network exists if a saccade was performed between those two targets. The weight of the edge is given by the number of times that specific saccade was performed. Therefore, the saccade network can be written as the directed and weighted $N \times N$ adjacency matrix, $\mathbf{S}$, whose element $S_{ij}$ denotes the weight of the edge between node $i$ and node $j$. Edges with zero weight signify no connection. 

We seek to identify unique saccade patterns across trials, which will in turn allow us to investigate how performance strategies might evolve throughout the task\cite{desrochers2010optimal}. We refer to the unique saccade pattern performed during a given trial as the trial representative saccade pattern (TRSP). To identify the TRSP, we utilize the saccade network of a given trial and identify all the cycles in the network. In graph theory, a cycle is defined as a series of edges that allows for a node to be reachable from itself. Therefore, a cycle in the saccade network is a loop that starts and ends on the same target. This cycle can be represented as a binary \textit{cycle matrix}, $\mathbf{L}$, of size $N \times N$; a given element of $\mathbf{L}$ is set to one if the corresponding edge is a part of the cycle. We take the dot product of the cycle matrix $\mathbf{L}$ and the saccade network $\mathbf{S}$, and refer to the resulting matrix as $\mathbf{L}'$. For a given saccade network, multiple cycles can exist, and therefore also multiple $\mathbf{L}'$s. We define the TRSP to be the cycle with the greatest element-wise sum of all weights in its $\mathbf{L}'$, which intuitively is the cycle composed of the most common point-to-point saccades. 

To ensure that the TRSP was not merely an outlier cyclic path performed by the monkey, we calculated the sum of edge weights corresponding to each cyclic saccadic path and assessed how much larger the TRSP-derived sum was compared to that of the remaining cyclic paths visited. For Monkey G, the averaged (across all trials and sessions) TRSP was 1 standard deviation away from the mean of weighted edge sums of all other cyclic paths (min-max standard deviations range: [0.45, 2.78]); for Monkey Y, the averaged TRSP was 1.01 standard deviation away from the mean of weighted edge sums of all other cyclic paths (min-max standard deviations range: [0.38, 3.06]). These results highlight two points: (i) the TRSP across each trial was not an outlier cyclic path performed by the monkey (when defining an outlier as a data point that is 4 standard deviations away from the mean), and (ii) the TRSP patterns of both monkeys were very similar to each other (as shown by the aforementioned, highly similar min, max, and mean ranges). For an intuitive graphical depiction of this process, see \textbf{Supplementary Figure 1}.

To characterize the similarity between trial saccade sequences, we first considered the challenge of comparing two saccade patterns. Many statistics exist for comparing graphs, but it is often unclear which is most appropriate in a given context \cite{wills2019metrics}. To circumvent this issue, it is useful to consider the cyclic path between nodes in a single TRSP to be a 2-D polygon drawn on the 3$\times$3 grid of equally sized and spaced circles presented during the task. Every saccade that is a part of the pattern is represented as a straight line between two centers of circles on the grid. Each line can then be discretized into small segments, allowing it to be summarized as the set of 100 (x,y) coordinates of segment centers. In other words, each saccade pattern can be summarized by the set of $n$ points, $P$:
        \begin{equation}
            P=\{\left(\mathrm{x}_i,\mathrm{y}_i\right)\mid\left(\mathrm{x}_i,\mathrm{y}_i\right)\in R^2\}, \mathrm{for}\;i = 1,\ldots n .
        \end{equation}
which finely samples the lines composing the \textit{saccade polygon}.

While a set of points is a simpler representation than a graph, it remains difficult to compare these point sets in a manner that accounts for the original geometry. To address this difficulty, we first calculate the centroid of the saccade polygon, $C$, and then we calculate the Euclidean distance between $C$ and every point in $P$:
        \begin{equation}
             D_i=\sqrt{\left(P\left(\mathrm{x}_i\right)-C\left(\mathrm{x}_i\right)\right)^2+\left(P\left(\mathrm{y}_i\right)-C\left(\mathrm{y}_i\right)\right)^2}
        \end{equation}
Indeed, we found the centroid measure to be robust to noise (\textbf{Supplementary Analysis 1}) and to exhibit a 1:1 relationship with its corresponding TRSP (\textbf{Supplementary Analysis 2}). Then $D$ is a 1-D \textit{saccade waveform} that parsimoniously represents the saccade polygon while maintaining geometric information. This 1D interpolation represents a standard approach to describing polygons that yields a continuous function. To ensure comparability across polygons and that all saccade waveforms are being compared to the same number of data points, we interpolate each D to I = 600 points. Using this approach, we could include a large number of data points that would generate a continuous function and thereby accurately capture differences across saccadic waveforms. To assess the robustness of our results to other choices of $I$, we also used a range of different $I$s (range: 400 to 700) and re-calculated the variables of interest: average similarity factor, average complexity factor, and cluster label entropy, described below. The recomputed variables were statistically highly similar (Pearson's $\rho$ > 0.7; $p < 10^{-6}$, across all comparisons) to the ones obtained when using $I = 600$ points, indicating that our choice was a representative one. \par

The fact that the saccade waveform can be thought of as a time-series representation of a saccade pattern informs our measure of similarity. Importantly, we wish our measure to be invariant to rotation and reflection, such that two polygons rotated by 90 degrees from one another or two polygons which are direct mirror images of each other are correctly determined to be the same. Therefore, rather than simply calculating the Euclidean distance between two saccade waveforms, we instead calculated the Euclidean distances between one saccade waveform and a series of circularly shifted versions of the second saccade waveform. A circular shift is a mathematical operation where a vector is rearranged such that the last element is moved to the first position and all other elements are shifted forward by one. By performing this operation $l$ times, it is possible to shift the last $l$ values to the front of the vector and all other values forward by $l$ positions. 

Accordingly, we create a set of circularly shifted saccade waveforms that represent rotations of the original saccade sequence by different angles as well as rotations of the mirror image of the original saccade sequence. The mirror image of the saccade sequence can be represented by flipping the saccade waveform from left to right (see \textbf{Supplementary Figure 7}). We write the angle of rotation as $\alpha=360(\frac{l}{I})$. For each pair of saccade sequences, a two-step approach is used to quantify their dissimilarity. First, for each pair we calculated the Euclidean distances between one saccade waveform and the circular shifts of a second saccade waveform (including its mirror image representation) in intervals of $\alpha=6$ degrees such that $l\approx10$. From this set of calculations we identify the circular shift which resulted in the smallest Euclidean distance and denote its index as $i_{min}$. Next, we repeated the above process but now performed shifts of size $l=1$ such that $\alpha\approx0.58$. These fine-grained secondary calculations included only circular shifts between $i_{min}-1$ and $i_{min}+1$. The dissimilarity factor (DF) between the two saccades was taken to be the minimum of the calculated distances in the second step; saccades with large distances between them are more dissimilar.

To summarize the saccade pattern similarity between two adjacent trials within a task session, we defined the \textit{SF}. For each session $s$ containing $T_s$ trials, the saccade dissimilarity was calculated between saccade patterns from pairs of consecutive trials as described in the previous section. Each value was then converted into a measure of similarity as follows: $SF = 1 - \frac{DF}{DF_{max}}$, where $DF_{max}$ is the maximum DF observed out of all trials from both monkeys. For each session the \textit{ASF} was then given by the average of all SFs calculated for that session. See \textbf{Supplementary Figure 8a} for the ASF as a function of session for both monkeys.

To characterize the complexity of each saccade pattern, we defined the \textit{CF}. We operationalized the notion of complexity as the fractal dimension\cite{mandelbrot1983fractal,iannaccone1996fractal}, which has proven useful in the study of many other biological\cite{smith1996fractal,liu2003fractal} and network systems\cite{song2005self}. For each TRSP, we first constructed a binary image of its polygon representation, where the background was black and the saccade pattern outline was white. We then applied the box-counting method to this image to compute the fractal dimension\cite{li2009improved}. Due to the nature of this method, two identical patterns rotated by 90 degrees from one another would result in different fractal dimension values. Therefore, the final fractal dimension value for each trial pattern was taken to be the minimum calculated from the set of all possible rotations of the pattern and its mirror image at intervals of 90 degrees. For each session, $s$, containing $T_s$ trials, the \textit{ACF} was given by the average fractal dimension of all trial representative saccade patterns in that session.  See \textbf{Supplementary Figure 8b} for the ACF as a function of session for both monkeys.

To quantify the extent to which the monkey selects saccade patterns from trial to trial in an ordered fashion, we defined the \textit{CLE} metric. More precisely, our goal was to determine whether the monkey was randomly performing various patterns or selectively repeating only a few unique ones. We began by using the MATLAB function \textit{linkage()} to perform agglomerative clustering on the unique saccade waveforms from all the trials of an individual monkey\cite{rokach2005clustering}. The algorithm outputs a hierarchical, binary cluster tree, also known as a dendrogram, based on an input of a $T \times T$ distance matrix, where the $ij$-th element gives the $DF$ between the saccade pattern of trial $i$ and the saccade pattern of trial $j$. The height of a link between two objects in the dendrogram directly denotes the distance between those two objects in the data. \par

It is important to note that the dendrogram itself does not indicate the optimal number of clusters that the data should be split into; rather, it demonstrates the order in which objects should be clustered. However, there is a way to identify the natural divisions of the data into distinct clusters using the derived cluster tree. The inconsistency coefficient metric is used to compare the height of a link in a cluster tree to heights of all the other links underneath it in the tree. If the difference is dramatic, it signifies that that newly formed group consists of linking two highly distinct objects. Imposing a threshold on the inconsistency coefficient during clustering captures more natural divisions in the data rather than arbitrarily setting a maximum number of possible clusters. \par

Accordingly, using the MATLAB function \textit{cluster()} we constructed clusters from the hierarchical cluster tree using a range of inconsistency coefficient values (0.1-1.5 in intervals of 0.05) as a threshold criterion, and then calculated the average within cluster sum-of-squares. Using the elbow-method and the calculated within cluster sum-of-squares, we selected an inconsistency coefficient of 0.95 to be the optimal threshold criterion for clustering. This choice resulted in a total of 136 clusters being identified for Monkey G (out of a total of 276 distinct paths) and 346 for Monkey Y (out of a total of 801 distinct paths). See \textbf{Supplementary Figures 2 and 3} for a detailed listing of cluster patterns of both monkeys.\par

After coarse-graining the data by identifying saccade clusters across trials and sessions, we next turned to assessing whether the monkey transitioned between saccades of the same cluster or of different clusters, and to what degree. We began by identifying the representative saccade sequence of each discovered cluster by finding the TRSP that had the minimum sum of the DF when compared to all other within-cluster patterns. For each session, we then calculated the \textit{CLE} as Shannon's information entropy of a given session's vector of trial cluster labels.  See \textbf{Supplementary Figure 8c} for the saccade cluster entropy as a function of session for both monkeys.

\subsection*{Channel Firing Rates} 
\label{subsec:methods:fr}

Information about neuronal activity was not available for all channels during each session; some channels, which we refer to as non-viable channels, showed no activity across all task trials. All non-viable channels were discarded prior to data analysis. Monkey G contained a total of 59 viable channels and Monkey Y contained a total of 64 viable channels. On average, a given free-scanning task session contained 23 active channels for Monkey G and 11 active channels for Monkey Y. While biologically expected, this variation in channel availability across sessions can adversely affect estimates of effective connectivity (see next section). For example, if we were to compute the effective connectivity from the activity of all channels across all trials at the same time, we could obtain spurious results due to the incomplete sampling across trials and time. \par

To mitigate potential biases due to variable channel availability, we estimate the effective connectivity in each session separately. Every session contains $N_{CH}$ channels from which a signal is available. The signal from each channel is in the form of a spike train, or a vector of ones signifying neuronal activation, and the time at which each activation occurred. All spikes from the full duration of a trial were used when inferring effective connectivity. For analysis concerned with identifying the relationship between behavior and control energy, we focused solely on spikes that occurred after the task grid was presented to the monkeys ($t_{targets\;on}$) and before the task grid disappeared signifying success ($t_{targets\;off}$). This window of time is referred to as the scanning period ($t_{sp}$). 

The firing rate $r$ of a single channel is given by the number of spikes per second.  Calculation of the fire rate of all channels during an individual trial then results in a $1 \times N_{CH}$ vector that represents the activation state of the channel network during that trial. We will refer to this column vector as the neural state $x_{t}$:
        \begin{equation}
            x_t=\left[\begin{matrix}r_1\\r_2\\\vdots\\r_{N_{CH}}\\\end{matrix}\right]\mathrm{, ~~where~t=}1,\ldots ,T_s
        \end{equation}
State vectors $x_t$ are calculated for each trial, across all sessions, and for each monkey. 
        
\subsection*{Inferring Effective Connectivity} \label{subsec:methods:ec}

Effective connectivity provides information that differs from both functional connectivity and structural connectivity\cite{friston2011functional}. For nearly three decades, effective connectivity has been ``understood as the experiment and time-dependent, simplest possible circuit diagram that would replicate the observed timing relationships between the recorded neurons''\cite{aertsen1991dynamics}. Effective connectivity moves beyond the purely correlational nature of canonical functional connectivity and provides a means of statistically inferring activity flow across brain regions\cite{friston2011functional}. Prior studies from our group and others have previously used effective connectivity in combination with network control theory to identify neural correlates of brain-computer interface learning\cite{stiso_bci} and points of intervention for seizure suppression\cite{scheid2020pnas}, highlighting the feasibility and utility of this approach. In addition, it is important to note that the control of functional connectivity networks necessitates the use of different methods than the control of structural or effective networks\cite{FC_control2019}.

The pattern of effective connectivity among many units can be usefully represented as a network, composed of nodes (neural units) and edges (effective connections) derived from node activity. Here, we construct such a network for the set of electrode channels used to record neuronal activity during trials of the free-scanning task. We chose transfer entropy as the method to estimate effective connectivity\cite{vicente2011transfer}, although we acknowledge that other methods exist and could similarly prove useful in the study of habit learning. Transfer entropy, while unable to infer causal influence between time-series, represents a closer approximation to this goal compared to purely associative methods like Pearson’s or Spearman’s correlation, which are commonly employed for defining functional connectivity. 

Mathematically, the (Shannon) transfer entropy was defined as follows:

\begin{equation}
TE_{Y \rightarrow X} = \sum_{i,j}p(i_{t+1},i_{t}^{(k)},j_{t}^{(l)}) \cdot log(\frac{p(i_{t+1}|i_{t}^{(k)},j_{t}^{(l)})}{p(i_{t+1}|i_{t}^{(k)})})
\end{equation}

\noindent{where $i_{t}$ denotes the status of signal time-series $X$ at time $t$ and \revtwo{reflects the average firing rate associated with a given trial}; $j_{t}$ denotes the status of signal time-series $Y$ at time $t$; $i_{t}^{(k)} = (i_{t},...,i_{t-k+1})$ and $j_{t}^{(l)} = (j_{t},...,j_{t-l+1})$ with $k$ and $l$ denoting the order of the transfer entropy and representing the past spiking history of the signals $X$ and $Y$, respectively; $p$ denotes probability; and the vertical bar denotes conditional probability. Overall, $TE_{Y \rightarrow X}$ quantifies the information flow from signal time-series $Y$ to $X$, and is based on the assumption that a time-series $Y$ can better predict the behavior of signal time-series $X$, if the combined past spiking histories of $Y$ and $X$ are more informative in predicting the future spiking history of time-series $X$ than the past history of $X$ alone\cite{vicente2011transfer,ito2011transferentropy}.}

For each session, we computed the transfer entropy between all pairs of viable-channels from the set of z-scored state vectors $x_t$ of size $1 \times N_{CH}$ to obtain an $N_{CH} \times N_{CH}$ directed, asymmetrical effective connectivity matrix whose diagonal elements are set to zero. All calculations of transfer entropy were performed using the \textit{calc\_te()} function from the \textit{RTransferEntropy} package in R. \par

Rather than utilizing a pre-defined temporal window when estimating transfer entropy, we calculated the overall transfer entropy for each session. For completeness, we gathered all session effective connectivity matrices into the 3-D matrix $\mathbf{M}$ whose element $M_{k,i,j}$ represents the effective connectivity between channels $i$ and $j$, derived from the $k^{th}$ session. From the individual session effective connectivity matrices, we calculated the overall effective connectivity matrix, $\mathbf{\mathcal{M}}$, whose element $\mathbf{\mathcal{M}}_{i,j}$ is given by the average of the set of the non-zero connection strengths between channels $i$ and $j$ derived from only the sessions in which both channels were available. Accordingly, $\mathbf{\mathcal{M}}$ is a square directed and asymmetrical matrix of size $N_{TC} \times N_{TC}$, where $N_{TC}$ is the total amount of available channels across all sessions for an individual monkey.

Lastly, to ensure that the generated effective connectivity matrices are specific to the monkeys' functional patterns of activation rather than noise, we compared the nodal degree distribution of our observed transfer entropy-derived effective connectivity matrix to the nodal degree distribution of a randomly generated set of effective connectivity matrices. Each one of these null connectivity matrices was generated by (i) randomly shuffling the average trial-based firing rate patterns corresponding to each channel ($n$ = 1000 permutations), across channels, (ii) calculating the new transfer entropy based on these random assignments, and (iii) generating null effective connectivity matrices using these newly-derived transfer entropies. In both monkeys, the nodal degree distribution corresponding to their observed effective connectivity matrix was significantly higher than the nodal degree distribution corresponding to the null matrices (two-sample Kolmogorov-Smirnov test -- Monkey G: $ks_{stat}$ = 0.3051, $p$ = 0.0062; Monkey Y: $ks_{stat}$ = 0.5, $p = 1.01\times{10^{-7}}$).

\subsection*{Network Control Theory} \label{subsec:methods:controlenergy}

To build an intuition for how we use network control theory to probe relations between neural circuit activity and behavior, we begin with a few preliminaries. We consider a non-linear dynamical system and linearize those dynamics about the system's current operating point\cite{kailath1980linear}. The dynamics of the resultant linear time invariant (LTI) system can be written as:
        \begin{equation}
            \dot{x}=\mathbf{A}x\left(t\right)+\mathbf{B}u\left(t\right)
        \end{equation}
where $N$ is the number of nodes, $\mathbf{A}$ is the $N \times N$ adjacency matrix,\\
$x\left(t\right) = \left[x_1\left(t\right),\allowbreak x_2\left(t\right),\allowbreak \ldots, \allowbreak x_N\left(t\right)\right]$ is the state of all network nodes at time t, $u\left(t\right)=\left[u_1\left(t\right),u_2\left(t\right),\ldots u_K\left(t\right)\right]$ gives the external control input for $K$ number of \textit{driver nodes} which receive external input in order to drive the state change of the network. In this work, all network nodes are set to be driver nodes ($K = N$) and as such $\mathbf{B}$ is the $N \times N$ identity matrix. 

The minimum control energy is defined to be the minimum amount of energy that a controller requires to drive an LTI system from some initial state to a target final state in a specified amount of time\cite{kailath1980linear}. If there exists an input vector $u\left(t\right)$ that can move the defined network from its initial state, $x_o$, to a state $x_f$ in time $t_f$, the energy expenditure is given by $E(t_f)=\int_{0}^{t_f} \left\lVert u(\tau)\right\rVert ^{2}d\tau$. Practically, we can calculate the minimum control energy to reach the target network state ($x_f$) from an initial state ($x_o$) as
        \begin{equation}
            E_{\mathrm{min}}\left(t_f\right)=\left(e^{\textbf{A}t_f}x_o-x_f\right)\textbf{W}_c^{-1}\left(t_f\right)\left(e^{\textbf{A}t_f}x_o-x_f\right)\ 
        \end{equation}
where $t_f$ is the time horizon and $\textbf{W}_c\left(t_f\right)$ is the controllability Gramian,
        \begin{equation}
            \textbf{W}_c\left(t_f\right)=\int_{0}^{t_f}{e^{\textbf{A}\tau}BB^{\top} e^{\textbf{A}^{\top}\tau}}d\tau
        \end{equation}
for the system. The numerical method used to solve the integral is MATLAB's native numerical ``integral'' command, and the time horizon is set to a value of 1 (units: $dt$) for all calculations in this analysis, as previously reported\cite{betzel2016optimally}. We provide references and a discussion on the pros and cons of this method in the Discussion.

Here we use this framework to compute the ACE required to move the neural circuit from firing rate state $x_t$ to firing rate state $x_{t+1}$ given the effective connectivity $\mathbf{A}_s = \mathbf{\mathcal{M}}_{i,j}$ for all channels in that session. Note that $x_t$ and $x_{t+1}$ are the firing rate states of two consecutive trials within a given session. Thus, we obtain an $E_{min}$ value for every consecutive trial pair. The ACE metric is then the average of $E_{min}$ across all state transitions in that session.

\subsection*{Control Energy \& Saccade Characteristics} \label{subsec:methods:acecorr}

We calculated Pearson correlation coefficients between the three saccade characteristic metrics (SF, CF, CLE) and the ACE. \revtwo{To control for variability in the number of recorded channels (NCH) across sessions, we regressed each variable of interest on NCH using a linear model. We then computed correlations between the resulting residuals, thereby assessing the relationship between variables independent of NCH.} To ensure that all correlations were specific to the control energy dynamics derived from the inferred overall effective connectivity matrix, $\mathbf{\mathcal{M}}$, we permuted that vector of ACE (number of elements = number of sessions considered for each monkey) 1000 times and calculated the Pearson correlation between the permuted ACE vector and the original saccade metric vector (SF, CF, CLE), for each permutation, to generate a null distribution of correlation coefficients. A one-tailed test was then used to determine the significance of the Pearson correlation coefficients calculated with the original ACE dynamics against their respective null distributions, at a significance level of $\alpha=\ 0.05$. 

We ran a simulation in order to demonstrate that directional tuning alone does not explain our reported results. Instead of using each Monkey’s actual firing rates obtained across each session, channel and trial, to generate its empirical effective connectivity matrix and ACE, we generated a ``null matrix'' of firing rates using only the monkeys’ saccadic behavior. This null matrix was then used to re-compute the monkeys’ simulated effective connectivity matrices and corresponding average control energies. The steps we took to accomplish this computation were:

For each monkey and for each session: 

1. We first created an array that contained all possible saccadic angles ranging from 0 to 360 degrees that could be taken when traversing across the 9 grid points. 

2.  We then created a new matrix $C \times T$, where $C$ corresponded to the number of simulated cells (set equal to the number of channels used for that Monkey) and $T$ to the number of trials. 

3.  For each simulated cell, we randomly selected one of the angles reported in step (1) as the cell’s preferred direction (i.e., direction tuning). For that angle, we designated an arbitrary firing rate as the cell’s maximum firing rate. The maximum firing rate was represented as the peak of a Gaussian distribution in which distance from the center along the $x$ axis reflected the closeness of each angle to the preferred direction\cite{bruce1985primate}. 

4.  Using this mapping between saccadic angles and firing rates, we computed the simulated cell’s average firing rate for each trial, using the Monkey’s actual saccades. Thus, for instance, if the Monkey had performed 10 saccadic movements during Trial 1, we extracted the corresponding cell-specific simulated firing rate for each of the 10 saccadic movements using the mapping described in step (3), and then averaged across all 10 firing rates to obtain the average firing rate of cell $C$ during trial $T$.

5.  The null matrix $C \times T$ was then used to recompute the Monkey’s simulated effective connectivity matrix and corresponding ACE, as reported in the manuscript. 

6.  Lastly, we repeated all main analyses reported in the text using the Monkey’s simulated ACE.

The results corresponding to the aforementioned simulation are reported in the ``Simulation Analysis'' section of the Results.

\subsection*{Network Region Lesion Analysis}

For both monkeys, electrode channels recorded bilaterally from regions of the caudate nucleus, frontal eye fields, and prefrontal cortex. In order to examine the extent to which specific nodes or edges contribute to the inferred overall effective connectivity matrix, we performed a virtual lesion analysis. Here, a lesion is operationalized by setting specific elements $\mathbf{\mathcal{M}}_{i,j}$ of the effective connectivity matrix to zero, thereby effectively eliminating the connection between the $i^{th}$ and $j^{th}$ nodes. Each monkey had a unique set of regions, $R=\{R_1,R_2,\ldots ,R_N\}$, from which the $N_{TC}$ channels were recording such that each channel was assigned only one region across the entire duration of the task. 

For each monkey, we performed two types of lesions. First, we lesioned all the edges between any two regions, $R_i$ and $R_j$, and we refer to this method as the \textit{inter-region edge knockout}. Second, we lesioned all edges belonging to the same region, and we refer to this method as the \textit{intra-region edge knockout}. Each lesion results in a knockout effective connectivity matrix which we denote as $\mathbf{\mathcal{M}}^{KO}$. Therefore, if performing an inter-region edge knockout between the caudate nucleus ($R_1$) and Brodmann Area 8 ($R_2$) then $\mathbf{\mathcal{M}}^{KO}$ would be the same as the original effective connectivity matrix, $\mathbf{\mathcal{M}}$, except that all elements that represent connections between $R_1$ and $R_2$ are set to zero. In the same way, if performing an intra-region edge knockout lesion of the caudate nucleus then $\mathbf{\mathcal{M}}^{KO}$ would be the same as $\mathbf{\mathcal{M}}$ but have all elements that represent connections of caudate nucleus nodes to other caudate nucleus nodes set to zero. \par

To determine the relevance of a connection for an energy-behavior correlation, we used two criteria. The first criterion was that the lesion resulted in a correlation value that was not significantly different ($p>0.05$) from that obtained using the original permutation null model (random permutations of the original ACE vector values). The obtained $p$-value is referred to as $p_{general}$ (\textbf{see Supplementary Figure 9b and 9d}). To assess this criterion, we calculate the knockout ACE metric, $ACE^{KO}$, for each lesion and use it to recompute the Pearson correlation values between $ACE^{KO}$ and the average saccade metrics. We test the significance of each knockout correlation value using a one-tailed test against the null distribution derived from calculations involving the original permutation null model. A knockout correlation that fails to prove significant from the above one-tailed test signifies that the inter- (or intra-) region edges knocked out are important to the inferred effective connectivity matrix and its relationship to task behavior. 

The second criterion for determining the relevance of a connection for an energy-behavior correlation was that the lesion-induced disruption of the observed correlation was specific to the lesion chosen, and not expected by lesioning the same number of randomly chosen edges. To assess this criterion, every knockout correlation value is compared to the original correlation value by calculating the absolute value of the difference between them, resulting in a correlation difference metric for each lesion. This metric quantifies how strongly the removal of the targeted edges alters the energy-behavior relationship, allowing us to determine whether specific connections are particularly influential in maintaining the observed correlation. 

To ensure that the difference in correlations is truly related to the lesion-specific edges, the results were tested using a null model. The null hypothesis states that, for a lesion consisting of $n$ specific edges $E^{KO}=\{e_1,e_2,\ldots,e_n\}$, the resulting correlation difference value is no different from the correlation difference value derived from knocking out $n$ randomly selected edges, $E^{null}=\{e_i\mid e_i\notin E^{KO}\}$. For every lesion, a null distribution of 1000 correlation values was created by randomly knocking out the same number of edges as the original lesion, ensuring that no edge overlapped with those in $E^{KO}$. All values were tested against their respective null distributions using a one-tailed test with significance level $\alpha=\ 0.05$. The obtained p-value is referred to as $p_{lesion}$ (\textbf{see Supplementary Figure 9c and 9e}). Gray boxes shown in (\hyperref[fig6]{Figure 6B-C}) indicate edges which, when lesioned, did not result in a significant change in behavior-energy correlation compared to the null distribution derived from randomly lesioning edges.

Due to the small magnitude changes in behavior-energy correlations caused by inter-/intra-region virtual lesioning, we next sought to quantify the number of edge lesions that are required to completely disrupt an observed behavior-energy correlation. Accordingly, for each monkey the resilience of each behavior-control correlation was tested by performing increasingly large lesions and re-calculating the correlation values. The analysis started with 1-edge lesions and ended with lesions involving up to 95\% of all available edges. Each lesion was performed 100 times with random edges and the average changes in behavior-energy correlation are shown in \textbf{Supplementary Figure 10} for Monkey G and Monkey Y.

\section*{Data Availability} The data that support the findings of this study are available from the corresponding authors upon reasonable request.

\section*{Code Availability} Publicly available MATLAB code for implementing the analyses described in the manuscript can be found here: https://github.com/kpszym/SaccadePatternExtraction.git. 

\section*{Acknowledgments} We thank Eli J. Cornblath, Pragya Srivastava, Christopher W. Lynn, Harang Ju, Sophia David, Jennifer A. Stiso, William Qian, and David M. Lydon-Staley for helpful comments on an earlier version of this manuscript. The work was primarily supported by an ARO MURI awarded to Bassett \& Graybiel (Grafton-W911NF-16-1-0474). The work was further supported by the John D. and Catherine T. MacArthur Foundation, the Alfred P. Sloan Foundation, the ISI Foundation, the Paul Allen Foundation, the Army Research Laboratory (W911NF-10-2-0022), the Army Research Office (W911NF-21-1-0328 to AMG), the National Institute of Mental Health (2-R01-DC-009209-11, R01-MH112847, R01-MH107235, R21-MH-106799, R21MH125010 and R01MH131615), the National Institute of Child Health and Human Development (1R01HD086888-01), National Institute of Neurological Disorders and Stroke (R01 NS099348, R01-NS025529), the National Eye Institute (R01-EY012848), the National Institutes of Health Ruth L. Kirschstein National Research Service Award (T32-EB020087, F32-AA030475), an NSF CAREER Award (BCS-2143656 to TMD), the National Science Foundation (NSF PHY-1554488, BCS-1631550, and IIS-1926757), and the Cornell Bethe/KIC/Wilkins postdoctoral fellowship, the Cornell Neurotech Mong Family Foundation postdoctoral fellowship, and the Eric and Wendy Schmidt AI in Science postdoctoral fellowship to JZK. The content is solely the responsibility of the authors and does not necessarily represent the official views of any of the funding agencies.

\section*{Author Contributions} D.S.B, T.M.D. and A.M.G. conceived of the study; T.M.D. and A.M.G. contributed data; J.K.B., P.F., K.S., J.Z.K. and F.P. carried out all analyses and interpreted the results; J.K.B., P.F., K.S., T.M.D. and D.S.B wrote the manuscript; all authors reviewed and edited the manuscript.

\section*{Competing Interests} The authors declare no competing interests.

\section*{Citation Diversity Statement} Recent work in several fields of science has identified a bias in citation practices such that papers from women and other minorities are under-cited relative to the number of such papers in the field\cite{mitchell_gendered_2013,dion_gendered_2018,caplar_quantitative_2017, maliniak_gender_2013,dworkin2020extent}. Here we sought to proactively consider choosing references that reflect the diversity of the field in thought, form of contribution, gender, and other factors. We obtained predicted gender of the first and last author of each reference by using databases that store the probability of a name being carried by a woman\cite{dworkin2020extent,zhou2020gender}. By this measure (and excluding self-citations to the first and last authors of our current paper), our references contain 10.6\% woman(first)/woman(last), 5.8\% man/woman, 19.2\% woman/man, 57.7\% man/man, and 6.73\% unknown categorization. This method is limited in that a) names, pronouns, and social media profiles used to construct the databases may not, in every case, be indicative of gender identity and b) it cannot account for intersex, non-binary, or transgender people. We look forward to future work that could help us to better understand how to support equitable practices in science.

\newpage
\printbibliography

\renewcommand\theequation{\Alph{section}\arabic{equation}} 
\counterwithin*{equation}{section} 
\renewcommand\thefigure{\Alph{section}\arabic{figure}} 
\counterwithin*{figure}{section} 
\renewcommand\thetable{\Alph{section}\arabic{table}} 
\counterwithin*{table}{section} 

\end{document}